\documentclass[11pt,a4paper]{article}
\pdfoutput=1
\usepackage[pdftex]{graphics}
\usepackage{jheppub}
\usepackage{amsmath,amssymb,amsfonts}
\usepackage{mathtools}
\usepackage{enumitem}
\usepackage{multirow}
\usepackage{array,booktabs}

\newcommand{\be}{\begin{eqnarray}}
\newcommand{\ee}{\end{eqnarray}}
\newcommand{\nn}{\nonumber}
\newcommand{\bn}{\begin{enumerate}}
\newcommand{\en}{\end{enumerate}}
\newcommand{\bl}{\begin{align}}
\newcommand{\el}{\end{align}}

\def\IC{\mathbb{C}}

\def\IP{\mathbb{P}}
\def\IR{\mathbb{R}}
\def\IZ{\mathbb{Z}}

\def\CL{{\cal L}}

\def\CN{{\cal N}}

\def\a{\alpha}
\def\b{\beta}

\def\e{\epsilon}

\def\l{\lambda}

\def\s{\sigma}

\def\L{\Lambda}
\def\thalf{{\textstyle \frac{1}{2}}}
\def\imp{\Longrightarrow}

\def\det{{\rm det}}

\def\jmath{{j}}

\renewcommand{\d}[1]{\mathinner{d#1}} % Defines the differential operator `d'
\title{Positroid Stratification of Orthogonal Grassmannian 
\\
and ABJM Amplitudes}

\author[a]{Joonho Kim,} 
\author[a,b,c,d]{Sangmin Lee\,} 

\affiliation[a]{Department of Physics and Astronomy, Seoul National University, Seoul 151-747, Korea}
\affiliation[b]{Center for Theoretical Physics, Seoul National University, Seoul 151-747, Korea}
\affiliation[c]{College of Liberal Studies, Seoul National University, Seoul 151-742, Korea}
\affiliation[d]{School of Physics, Korea Institute for Advanced Study, Seoul 130-722, Korea}

\abstract{
A novel understanding of scattering amplitudes in terms of 
on-shell diagrams and positive Grassmannian has been recently established 
for four dimensional Yang-Mills theories and three dimensional 
Chern-Simons theories of ABJM type. 
We give a detailed construction of the positroid stratification 
of orthogonal Grassmannian relevant for ABJM amplitudes.  
On-shell diagrams are classified by pairing of external particles. 
We introduce a combinatorial aid called `OG tableaux' and 
map each equivalence class of on-shell diagrams to a unique tableau. 
The on-shell diagrams related to each other through BCFW bridging 
are naturally grouped by the OG tableaux. Introducing suitably ordered 
BCFW bridges and positive coordinates, we construct 
the complete coordinate charts to cover the entire positive orthogonal 
Grassmannian for arbitrary number of external particles. 
The graded counting of OG tableaux suggests that 
the positive orthogonal Grassmannian constitutes a combinatorial polytope. 
}

\preprint{SNUTP-14001, KIAS-P14041}

\begin{document}

\maketitle

%\newpage

\section{Introduction and Conclusion}
The study of scattering amplitudes in gauge theories 
has gone through many stages of exciting developments 
(see \cite{Elvang:2013cua} for an up-to-date review). 
A common theme in many recent breakthroughs is reformulation of the gauge theory 
in such a way to uncover hidden structures that are hardly visible 
in the traditional approach. 

The Grassmannian integral \cite{ArkaniHamed:2009dn} is a relatively new reformulation 
notably far removed from the traditional approach. Tree amplitudes and loop integrands of a planar gauge theory are produced from a contour integral over the Grassmannian G$(k,n)$, 
where $n$ is the total number of external particles 
and $k$ is the number of external particles with negative helicity. The geometric structure makes the conformal symmetry and dual conformal symmetry completely manifest while relegating locality and unitarity to an emergent property. 
Although the Grassmannian integral inherits some crucial features from its predecessors 
such as the twistor string theory \cite{Witten:2003nn,Roiban:2004yf} and the BCFW recursion relation \cite{Britto:2004ap,Britto:2005fq}, 
its physical origin had remained elusive for years.

In a remarkable paper \cite{ArkaniHamed:2012nw}, 
Arkani-Hamed et al. introduced the notion of `on-shell diagrams' which 
in a sense provides a microscopic structure underlying the Grassmannian integral. The vertices of on-shell diagrams 
are the gauge invariant three-point amplitudes with $(++-)$ or $(+--)$ helicity assignments. They showed how to rewrite the vertices in 
terms of integrals over G$(1,3)$ or G$(2,3)$. By integrating out internal lines, 
they combine small Grassmannian integrals to form a big integral over G$(k,n)$ for 
arbitrary $k$ and $n$. The BCFW deformation of momenta appears naturally 
in on-shell diagrams and `bridges' two external legs with a new internal line and 
two vertices. 
While probing the structure of Grassmannian integral at a deeper level, 
the authors of \cite{ArkaniHamed:2012nw} also encountered a number of modern topics in mathematics such as affine permutation, positive Grassmannian, positroid stratification, 
and cluster algebra.
They also noted that the on-shell diagram approach 
can be applied to any gauge theory in four dimensions, 
although it takes by far the simplest form for the $\CN=4$ super-Yang-Mills theory (SYM$_4$). 

In three dimensions, a class of $\CN=6$ superconformal Chern-Simons matter theories \cite{Aharony:2008ug,Hosomichi:2008jd,Hosomichi:2008jb,Bagger:2008se,Schnabl:2008wj}, widely known as ABJM theory, serves as a main testing ground 
for novel methods on scattering amplitudes. An initial 
step toward the on-shell diagram approach to ABJM theory was already taken in \cite{ArkaniHamed:2012nw}, based on the an integral over the 
orthogonal Grassmannian OG$(k,2k)$ \cite{Lee:2010du}. 
The on-shell diagrams of ABJM theory consist of a unique quartic vertex and an internal line. The permutation governing the planar diagrams is a complete pairing of the $2k$ external particles. Two diagrams sharing the same permutation are equivalent to each other up to Yang-Baxter equivalence moves. 
The BCFW bridge connects two external legs by creating an extra vertex between them.

Recently, Huang and Wen \cite{Huang:2013owa} further studied the on-shell diagrams for ABJM amplitudes.
They refined the on-shell diagrams with some sign factors to account for the two disjoint `branches' of orthogonal Grassmannian.   
This is crucial since the full tree amplitudes receive contributions from both branches, as noted earlier in \cite{Gang:2010gy}.
By rewriting the BCFW recursion relation for ABJM theory \cite{Gang:2010gy} 
in terms of on-shell diagrams, they constructed on-shell diagram representation of all tree-level amplitudes. 
They also introduced a set of coordinates in which 
all consecutive minors take a simple form. 
Finally, they defined the notion of positive orthogonal Grassmannian. 
In contrast to the ordinary Grassmannian G$(k,n)$, it is slightly non-trivial 
to define reality conditions on OG$(k,2k)$. 

The aim of this paper is to cover a topic that was notably missing 
in \cite{ArkaniHamed:2012nw,Huang:2013owa}.  
Given a permutation, 
it is desirable to select a particular representative 
of the equivalence class of on-shell diagrams, and assign canonically positive coordinates to it. 
The construction of canonical coordinates elucidates 
how the Grassmannian integral is decomposed into 
a series of BCFW bridging. 
The canonical coordinates also enjoy several nice properties; 
for instance, the measure of the Grassmannian integral 
takes a simple `$d$(log)' form, and the boundaries 
of the positive orthogonal Grassmannian become zero loci 
of the coordinates. 
This problem of assigning canonical coordinates was solved completely in four dimensions in \cite{ArkaniHamed:2012nw} using a relation to the mathematics of `positroid stratification' \cite{Postnikov:2006kva,Knutson:2011}. We will solve the problem in three dimensions by introducing a combinatorial device 
similar to those of \cite{Postnikov:2006kva}, which we call `OG tableaux'. 
To our knowledge, mathematical studies of positive orthogonal Grassmannian remain incomplete and not readily accessible to physicists. See, e.g., \cite{Lam:2007} for a related work.

The rest of this paper is organized as follows.
Sec.~\ref{sec:on-shell} is a brief 
review of what is known for on-shell diagrams for ABJM theory 
from earlier works \cite{ArkaniHamed:2012nw,Huang:2013owa}. 
The elementary 4-particle vertex can be understood both as a BCFW bridge 
and an integral over OG$(2,4)$. Integration over internal lines 
`amalgamates' copies of OG$(2,4)$ and builds up bigger OG$(k,2k)$. 
In Sec.~\ref{sec:OG}, we take a closer look at the geometry of orthogonal Grassmannian. We examine real slices of OG$(k,2k)$ and discuss how to define positivity on a real slice. The real version of the BCFW bridge OG$(2,4)$ 
and amalgamation 
naturally suggests how to introduce manifestly real coordinates on 
the orthogonal Grassmannian. 
Sec.~\ref{sec:positroid} reports the main results of this paper.  
We begin with mapping the on-shell diagrams to what we call `OG tableaux'. 
The tableau notation serves several purposes. 
It fixes the ambiguity coming from Yang-Baxter equivalence relation 
in a canonical way. That helps us to count distinct on-shell diagrams, 
which are interpreted geometrically as cells of the positive Grassmannian. 
Moreover, 
the OG tableaux can be used to assign `canonical positive coordinates' for all on-shell diagrams. 
The integration measure in the canonical coordinates is a product 
of simple $d\log$ factors. 
The OG tableaux also help us to study mathematics of OG$_+(k,2k)$. The positive Grassmannian G$_+(k,n)$ 
is known to form a combinatorial polytope called `Eulerian poset' \cite{Williams:2007}. 
The graded counting of OG tableaux suggests that the positive orthogonal Grassmannian OG$_+(k,2k)$ similarly defines an Eulerian poset at each $k$. Geometrically, the graded counting hints 
at the possibility that OG$_+(k,2k)$ 
%regarded as a smooth manifold with boundary, 
may have a topology of a ball. We verify 
this conjecture for $k=2,3$ and leave the generalization to higher $k$ as an open problem.

This paper focuses on a formal aspect of orthogonal Grassmannian 
and makes little direct contributions to ABJM amplitudes. However, the results of this paper clearly opens up a few directions of further research. 
Here, we list three prominent possibilities. 

First, the issue of Yangian symmetry, which unites the ordinary and dual superconformal symmetries, could be revisited. 
While there are strong evidences for the Yangian symmetry 
of ABJM amplitudes \cite{Bargheer:2010hn,Lee:2010du,Gang:2010gy,Huang:2010rn,Huang:2010qy}, a formulation with manifest dual superconformal symmetry 
has not been found. For SYM$_4$, such a dual formulation was found earlier \cite{Mason:2009qx,ArkaniHamed:2009vw}
 and laid a foundation for further discoveries such as the `amplituhedron' \cite{Arkani-Hamed:2013jha}.
In \cite{ArkaniHamed:2012nw}, the Yangian symmetry was interpreted 
geometrically as diffeomorphisms which leave the measure on G$_+(k,n)$ invariant. A similar interpretation for OG$_+(k,2k)$, if possible, 
would shed new light on the Yangian symmetry of ABJM amplitudes. 

Second, our work could be related to twistor string models for 
ABJM amplitudes. For tree-level amplitudes of SYM$_4$, the equivalence 
between the twistor string formula \cite{Roiban:2004yf} and the Grassmannian formula \cite{ArkaniHamed:2009dn} was established in \cite{Bourjaily:2010kw}. Simply put, the derivation consists of three steps: specifying the integration contour of the Grassmannian integral, deforming its integrand with no loss of residues, 
and integrating out some variables. Along the same line of reasoning, 
a twistor string formula was proposed in \cite{Huang:2012vt} (and recently rederived from different viewpoints in \cite{Cachazo:2013iaa,Engelund:2014sqa}), 
but the derivation was less solid due to limited understanding of 
the integration contours. The canonical coordinates defined in this paper 
could be useful in finding a refined derivation comparable to that of \cite{Bourjaily:2010kw}. 

Third, the formal structure of the positive orthogonal Grassmannian could be probed at a deeper level. To define the canonical coordinate system, we suppressed the Yang-Baxter equivalence moves in a particular `frame'. If we move to another frame, 
the new coordinates would be related to the old ones in a non-trivial way. 
For SYM$_4$, the coordinate transformation has an intriguing connection 
to the mathematics of cluster algebra \cite{cluster,fock}. It would be interesting to figure out the ABJM counterpart of the story. 
See a recent work by Huang, Wen and Xie \cite{Huang:2014xza} for a related discussion. 

We conclude this introduction with some shorthand notations. 
We will write (P)OG for (positive) orthogonal Grassmannian, 
and abbreviate OG$(k,2k)$ and OG$_+(k,2k)$ by OG$_k$ and POG$_k$, 
respectively.

\paragraph{Note added} 
While an early version of this paper was being revised for publication, we received a preprint by Lam \cite{Lam} 
which rigorously prove a mathematical theorem stating that POG$_k$ defines an Eulerian poset for all $k$.

%\newpage

\section{On-shell Diagrams for ABJM Amplitudes \label{sec:on-shell}}

In this section, we review some salient features of the on-shell diagrams for ABJM amplitudes \cite{ArkaniHamed:2012nw,Huang:2013owa}. We first recall the kinematics 
of the ABJM amplitudes and the definition of the OG integral 
to set the stage for the on-shell diagram. 
After introducing the fundamental building blocks, a quartic vertex and an internal line, 
we examine two ways to construct the most general on-shell diagrams. 
One is BCFW bridging, which enables us to add a vertex one at a time to a given diagram. The other is amalgamation, which 
merges two diagrams into a larger one by integrating over an internal line. For both methods, we explain how the diagrammatics is reflected in the OG integral, thereby making 
the microscopic decomposition of OG integral manifest. 
Finally, we comment briefly on the `Yang-Baxter' equivalence relation for different diagrams corresponding to the same amplitude as well as the bubble diagrams.

\subsection{Elements of on-shell diagrammatics}
\label{sub:kinematics}

\paragraph{Kinematics}
The ABJM theory is a Chern-Simons-matter gauge theory 
in three dimensions with $\CN=6$ superconformal symmetry. 
The symmetry group is OSp$(6|4)$ whose bosonic part 
contains Sp$(4,\IR) \simeq {\rm Spin}(2,3)$ conformal symmetry 
and SO(6) $R$-symmetry. As shown in \cite{Bargheer:2010hn}, the OSp$(6|4)$ symmetry 
becomes manifest if we work in a superspace $\Lambda = (\lambda^\alpha;\eta^I) \in \IC^{2|3}$. 

In this representation, the superconformal generators come in three types:
\begin{align}
\L \frac{\partial}{\partial \L}
\,,
\;\;\;
\L\L \,,
\;\;\;
\frac{\partial^2}{\partial \L\partial \L} \,.
\label{suconf-gen}
\end{align}
The superspace notation will guarantee the invariance of the amplitudes under the $(\Lambda \partial/\partial \Lambda)$ generators. Let us decompose the $(\Lambda\Lambda)$ generators,
\begin{align}
p^{\a\b} = \l^\a \l^\b ,
\;\;\;
q^{\a I} = \l^\a \eta^I ,
\;\;\;
r^{IJ} = \eta^I \eta^J .
\end{align}
In a scattering process, the invariance under $p^{\a\b}$ and $q^{\a I}$ can be imposed by
the super-momentum conserving delta functions
\be
\delta^3(P) \delta^6(Q)
\;\;\; {\rm with} \;\;\;
P := \sum_i p^{\a\b}_i , \;\; Q := \sum_i q^{\a I}_i .
\ee
The $r^{IJ}$ invariance
introduces a coset ${\rm O}(2k-4)/{\rm U}(k-2)$
for the $2k$-point amplitude \cite{Bargheer:2010hn}. 
The coset structure was a precursor to the OG integral
for ABJM theory~\cite{Lee:2010du}.
Once the invariance under the $(\Lambda\Lambda)$ generators is confirmed, 
the invariance under the remaining $(\partial^2/\partial \Lambda\partial \Lambda)$ generators will follow from the superconformal inversion,  
which acts on $\Lambda$ as Fourier transformation ($\Lambda \leftrightarrow \partial/\partial \Lambda$).

The spinors $\l_i$ are contracted via the invariant tensor $\epsilon_{\a\b}$ ($\epsilon_{12}=-\epsilon^{12}=1$) of SL$(2,\IR) \simeq {\rm Spin}(1,2)$ Lorentz group, 
\begin{align}
\langle ij\rangle := \lambda_i^\alpha\lambda_{j\alpha}=\lambda_i^\alpha\epsilon_{\alpha\beta}\lambda_{j}^{\beta} \,.
\end{align}
The ABJM theory contains two matter multiplets with opposite gauge charge. 
The particle/anti-particle superfields take the form
\be
&&\Phi = \phi^4 + \eta^I \psi_I + \thalf \e_{IJK} \eta^I\eta^J \phi^K
+ \tfrac{1}{6} \e_{IJK} \eta^I \eta^J \eta^K \psi_4 \,,
\nn \\
&&\bar{\Phi} = \bar{\psi}^4 + \eta^I \bar{\phi}_I +
\thalf \e_{IJK} \eta^I\eta^J \bar{\psi}^K
+ \tfrac{1}{6} \e_{IJK} \eta^I \eta^J \eta^K \bar{\phi}_4 \,.
\label{sfield}
\ee
The color-ordered tree-level super-amplitudes, $A_{2k}(\L_1 , \cdots, \L_{2k})$, are functions of $\L_{i}$. Following the convention of \cite{Lee:2010du,Gang:2010gy}, we choose to associate $\L_{\rm odd/even}$ to $\bar{\Phi}$/$\Phi$ multiplet. 
As noted 
in \cite{Bargheer:2010hn}, the kinematics and the multiplet structure imply
the `$\Lambda$-parity', 
\begin{align}
A_{2k}( \Lambda_1, \cdots, - \Lambda_i , \cdots, \Lambda_{2k} )  = (-1)^{i} A_{2k}( \Lambda_1, \ldots , \Lambda_i , \cdots, \Lambda_{2k} )  \,.
\label{super-l-parity}
\end{align}
and symmetry under the cyclic shift  by two sites,
\begin{align}
A( \Lambda_1, \L_2, \cdots, \Lambda_{2k} ) 
= (-1)^k A(\L_3, \L_4, \cdots, \Lambda_{2k},\L_1,\L_2) \,.
\label{A-cyclic}
\end{align}

\paragraph{OG integral}
A central object of interest in this paper is 
the OG integral \cite{Lee:2010du}: 
\begin{align}
\CL_{2k}(\Lambda) =
\int  \frac{d^{k\times 2k} C}{{\rm vol}[{\rm GL}(k)]}\frac{\delta^{k(k+1)/2}(C\cdot  C^T)\, \delta^{2k|3k}(C\cdot \L)}
{M_1 M_2 \cdots M_{k-1} M_k}
\,.
\label{grass0}
\end{align}
The integration variable $C$ is a $(k\times 2k)$ matrix.
The dot products denote $(C\cdot C^T)_{mn} = C_{mi}C_{ni}$, $(C\cdot \L)_m = C_{mi}\L_i$.
The $i$-th consecutive minor $M_i$ of $C$ is defined by
\be
M_i = (C_i,C_{i+1},\cdots,C_{i+k-1}) = \e^{m_1 \cdots m_k} C_{m_1 (i)} C_{m_2 (i+1)} \cdots C_{m_k (i+k-1)} \,.
\ee
In \cite{Lee:2010du}, this formula was conjectured to reproduce the 
$2k$-point tree level amplitude upon a suitable choice of integration contour. The conjecture was verified up to $k=4$ in \cite{Gang:2010gy}.
%In the rest of this section, 
We will review the on-shell diagram approach to the OG integral \eqref{grass0}, initiated in \cite{ArkaniHamed:2012nw} and elaborated in \cite{Huang:2013owa}, in a way to facilitate the introduction of the positroid stratification %of orthogonal Grassmannian 
to be presented in Sec.~\ref{sec:positroid}.

\paragraph{Building blocks}
The on-shell diagrams for ABJM amplitudes 
are planar diagrams drawn on a disk with $2k$ boundary points 
representing cyclically ordered external particles.
Schematically, the building blocks of the diagrams 
take the following form \cite{ArkaniHamed:2012nw}:
\begin{align}
&\mbox{Internal line :} \hskip 0.8cm 
\raisebox{-0.5pt}{\includegraphics[width=1.4cm]{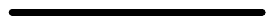}}
\hskip 0.8cm
 = \quad \int d^2 \lambda \,d^3 \eta  := \int d^{2|3} \Lambda 
 := \int ``d \Lambda " \,,
\label{def-internal0}
\\
&\mbox{Vertex :} \hskip 1.5cm
\raisebox{-30pt}{\includegraphics[width=2cm]{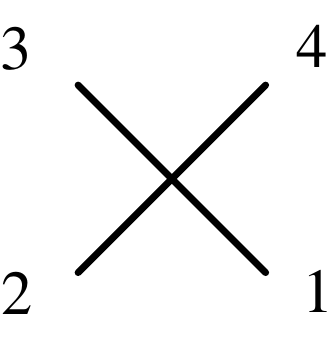}}
\hskip 0.5cm
= \quad \mathcal{A}_4(\Lambda_1, \Lambda_2, \Lambda_3,\Lambda_4) \,.
\label{def-vertex0}
\end{align}
The quartic vertex is precisely the 4-point tree amplitude first computed 
in \cite{Agarwal:2008pu} and reproduced from the OG integral in \cite{Lee:2010du,Gang:2010gy}.

%\paragraph{Cyclic Symmetry}
The graphical notations in \eqref{def-internal0} and \eqref{def-vertex0} 
are not fully well-defined as they stand. 
The 4-point amplitude $A_4$ does not have a $\IZ_4$ cyclic symmetry. Instead, it is odd under a cyclic shift by two sites, 
\begin{align}
A_4(1,2,3,4) = -A_4(3,4,1,2) \,.
\label{A4-cyclic}
\end{align}
Accounting for this symmetry, when the simpler notation is likely to cause confusion, we will use a refined notation for the vertex.
\begin{align}
\mbox{Vertex :} \hskip 2.1cm
\raisebox{-33pt}{\includegraphics[width=2cm]{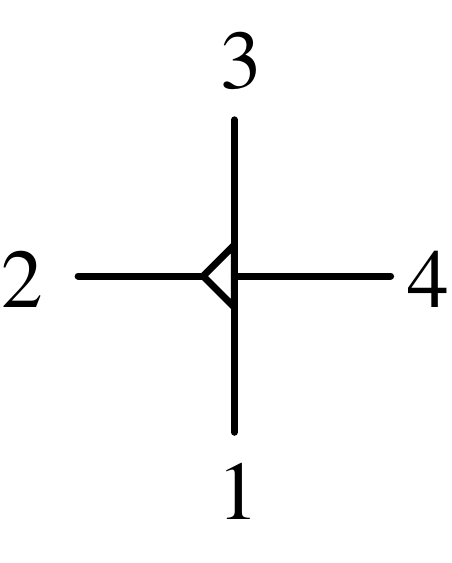}}
\hskip 1cm
= \quad
\frac{\delta^3(P) \delta^6(Q)}{\langle 12 \rangle \langle 23 \rangle}
\,.
\label{def-vertex1}
\end{align}

The internal line \eqref{def-internal0} means that 
two sub-diagrams, say $F(\Lambda)$ and $G(\Lambda)$, can be `glued' 
by an integral of the form, 
\begin{align}
\int d\Lambda \,F(\Lambda) G(-i\Lambda) =  \int d\Lambda_1 d\Lambda_2 \, \delta(i\Lambda_1 + \Lambda_2) F(\Lambda_1) G(\Lambda_2)  \,.
\label{def-internal1}
\end{align}
The factor of $i$ reflects momentum conservation between two particles: $\l\l+(i\l)(i\l)=0$. By convention, we will assign  
$\Lambda_1$ and $\Lambda_2$ in \eqref{def-internal1} to an odd $(\bar{\Phi})$
and an even $(\Phi)$ multiplet, respectively.

\subsection{BCFW bridge}
\paragraph{Vertex as BCFW bridge}
As shown in \cite{ArkaniHamed:2012nw,Huang:2013owa}, it is possible 
to build up complicated on-shell diagrams from 
a simpler one by adding vertices via `BCFW bridges'. 
In this subsection, we review how to interpret the elementary vertex \eqref{def-vertex1} as a BCFW bridge, and establish our convention 
for coordinates and sign factors. 

We recall from \cite{Gang:2010gy} that the BCFW deformation 
acts as an O$(2k,\IC)$ rotation on the kinematic variables $\Lambda_i$,  
which leaves the total super-momentum invariant. 
For the four-particle vertex, the total super-momentum is 
\begin{align}
P &= \l_1 \l_1 + \l_2 \l_2 + \l_3 \l_3 + \l_4 \l_4  = 0 \,,
\label{P4}
\\
Q &= \l_1 \eta_1 + \l_2 \eta_2 + \l_3 \eta_3 + \l_4 \eta_4  = 0 \,.
\end{align}
Moving $\l_3$ and $\l_4$ to the right-hand-side in \eqref{P4} 
and squaring, we find
\begin{align}
\langle 12 \rangle^2 = \langle 34 \rangle^2
\;\;
\imp
\;\; 
\langle 12 \rangle = \sigma \langle 34 \rangle 
\quad (\s = \pm 1) \,.
\end{align}
The sign factor $\sigma$ defines two `branches' 
of kinematic configuration. As we will see shortly, 
the same $\sigma$ will define two branches of OG$_2$. 

\begin{figure}[htbp]
	\centering
	\includegraphics[height=2.5cm]{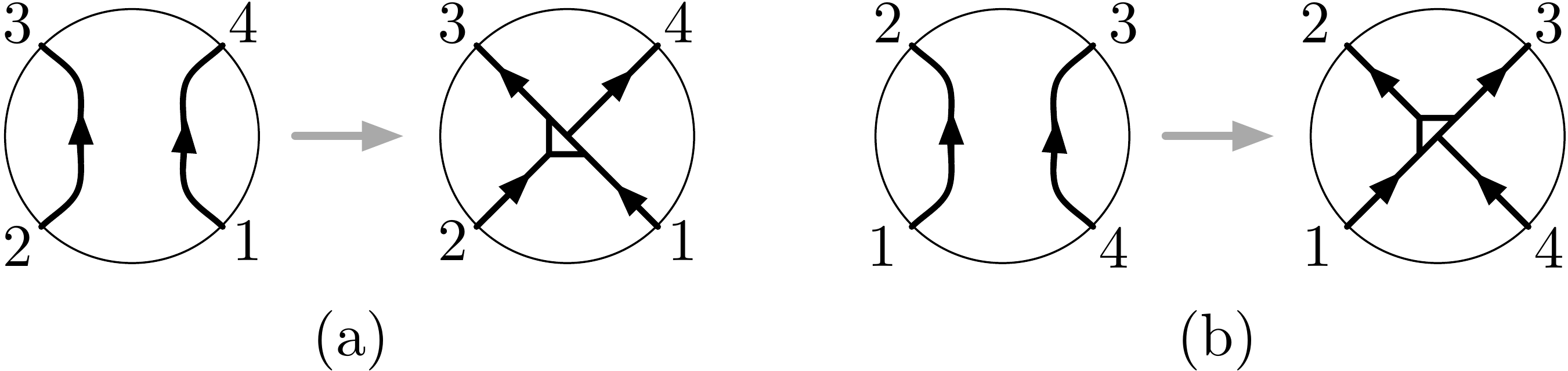}
\caption{4-point vertex as a BCFW bridge with two adjacent `sources' legs.}
	\label{fig:bcfw-vertex}
\end{figure}

To reveal the structure of the BCFW bridge, we begin with the vertex
\begin{align}
%A_4^{12\rightarrow 34} \equiv
A_4(\L_1,\L_2,\L_3, \L_4) = \frac{\delta(P)\delta(Q)}{\langle 12 \rangle \langle 23 \rangle } \,.
\label{eq:4ptint14a}
\end{align}
Inserting two identities,
\footnote{
Essentially the same computation was done in \cite{Bargheer:2012cp} 
with particular reality conditions on $\l_i$. Here, we leave 
$\l_i$ as complex variables, and treat the delta-functions 
as analytic functions.} 
\begin{align}
  &1 = -\langle34\rangle \int dc_{i3}\wedge dc_{i4} \, \delta^{2}(\lambda_i + c_{i3}  \lambda_3 + c_{i4}  \lambda_4) \qquad \text{for }i=1,2 \,,
\label{eq:iden1}
\end{align}
changes the momentum-conserving delta function into 
\begin{align}
  \delta^3(P) &= \delta^3 \left(\lambda_3\lambda_3 (1+c_{13}^2+c_{23}^2) + \lambda_4\lambda_4 (1+c_{14}^2+c_{24}^2) + (\lambda_3 \lambda_4 + \lambda_4 \lambda_3) (c_{13} c_{14}+c_{23} c_{24})\right) 
  \nonumber\\
    &= -\langle 34 \rangle^{-3} \, \delta(1+c_{13}^2 + c_{23}^2)\,\delta(1+c_{14}^2 + c_{24}^2)\,\delta (c_{13}c_{14} + c_{23}c_{24})\,.
\label{eq:P-14}
\end{align}
Taking the change of variables 
\begin{align}
\begin{pmatrix}
c_{13} & c_{14}  \\
c_{23} & c_{24}
\end{pmatrix}
= i
\begin{pmatrix}
r_3 \sin t_3 & \;\; r_4 \cos t_4  
\\
r_3 \cos t_3 & \;\; r_4 \sin t_4
\end{pmatrix} \,,
\label{eq:ch-14}
\end{align}
and integrating out $r_3$ and $r_4$, 
we find
  \begin{align}
    A_4
    &= -\frac{\delta^6 (Q)}{4i\langle 34 \rangle^3}\int \frac{\d{t_3} \wedge \d{t_4}}{\sin{t_4}\cos{(t_3 + t_4)}}\delta(\sin{(t_3 + t_4)}) 
    \nonumber \\
    &\hspace{2.3cm}\times\,\,\delta^{2}(\lambda_1 + i \sin t_3 \,\lambda_3 + i \cos t_4 \,\lambda_4) 
    \delta^{2}(\lambda_2 + i \cos t_3 \,\lambda_3 + i \sin t_4 \,\lambda_4)  \,\nonumber\\
    &= \frac{\delta^6 (Q)}{4i\langle 34 \rangle^3}\int \frac{\sigma \d{t_3}}{\sin{t_3}} \sum_{\sigma} \delta^{2}(\lambda_1 + i s_3 \,\lambda_3 + i \sigma c_3 \,\lambda_4) \; \delta^{2}(\lambda_2 + i c_3 \,\lambda_3 - i  \sigma s_3 \,\lambda_4) \,, 
  \end{align}
  where the first delta function is localized at $t_3 + t_4 = 0 \text{ or } \pi$. The fermionic delta function 
can be rearranged as 
 \begin{align}
    \delta^6(Q) &= -\langle12 \rangle^{-3}  \, \delta^3\left(\langle 12\rangle \eta_1 + \langle 32 \rangle \eta_3 + \langle42\rangle \eta_4\right) \, \delta^3\left(\langle 21 \rangle \eta_2 + \langle 31 \rangle \eta_3 + \langle 41 \rangle \eta_4 \right) \nonumber\\
    \label{eq:fermdeltaftn}
    &= \s \langle34\rangle^{3} \delta^{3}(\eta_1 + i \sin{t_3} \,\eta_3 + i \sigma \cos{t_3} \,\eta_4) \; \delta^{3}(\eta_2 + i \cos{t_3} \,\eta_3 - i \sigma \sin{t_3} \,\eta_4) \,.
  \end{align}
Collecting all the ingredients, we obtain
\begin{align}
  \label{eq:bcfw-4pt-vertex-12}
  A_4= \sum_{\s = \pm}\int \frac{\d{t}}{4i \sin{t}}
  \,\delta^{2|3}(\Lambda_1 + \L_4^{\s}(t))\,\delta^{2|3}(\Lambda_2 + \L_3^{\s}(t)) \,,
\end{align}
where
\begin{align}
\begin{pmatrix}
\L_3^{\s}(t) \\ \L_4^{\s}(t)
\end{pmatrix}
= 
\begin{pmatrix}
i \cos{t} & -i \sigma \sin{t}  \\
i \sin{t} & i \sigma \cos{t}  
\end{pmatrix}
\begin{pmatrix}
\L_3 \\ \L_4
\end{pmatrix} \,.
\label{eq:L12rot}
\end{align}
Pictorially, as shown in Fig.~\ref{fig:bcfw-vertex}(a), the result might be summarized as building a BCFW bridge between two `free propagators' $(14)$ and $(23)$. During the derivation, we chose two adjacent legs $\{1,2\}$ as `sources' and the other legs $\{3,4\}$ as `sinks'. 

\noindent
Other choices of sources and sinks are possible. 
Up to the cyclic symmetry \eqref{A4-cyclic}, 
the only other possibility for adjacent source legs is $\{1,4\}$, 
as depicted in Fig.~\ref{fig:bcfw-vertex}(b). 
With the branch parameter defined by $\langle 14 \rangle = \sigma \langle 23 \rangle$, $A_4$ becomes
\begin{align}
	\label{eq:bcfw-4pt-vertex-14}
	A_4= \sum_{\s = \pm}\int \frac{\d{t}}{4i \sin{t}} \,\delta^{2|3}(\Lambda_1 +\L_2^{\s}(t))\,\delta^{2|3}(\Lambda_4 + \L_3^{\s}(t)).
\end{align}
where
\begin{align}
\begin{pmatrix}
\L_2^{\s}(t) \\ \L_3^{\s}(t)
\end{pmatrix}
= 
\begin{pmatrix}
i\sigma \cos{t} & -i \sin{t}  \\
-i \sigma \sin{t} & -i \cos{t}  
\end{pmatrix}
\begin{pmatrix}
\L_2 \\ \L_3
\end{pmatrix} \,.
\label{eq:L14rot}
\end{align}

\noindent
We may also consider taking non-adjacent  
source legs, $\{1,3\}$ or $\{2,4\}$. They will not be used in subsequent sections, so we omit them here. Interested readers are referred to \cite{Huang:2013owa}. 

\begin{figure}[htbp]
	\centering
	\includegraphics[height=3cm]{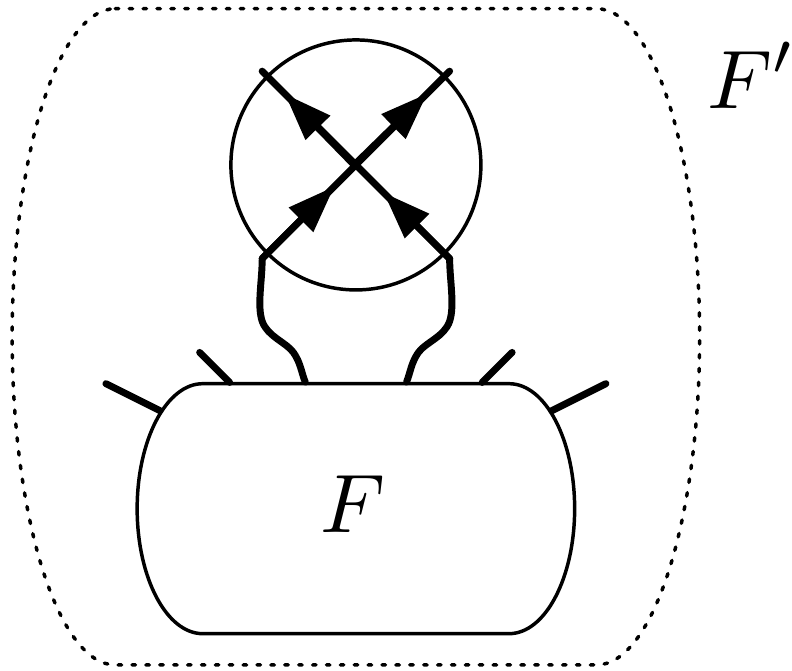}
\caption{Adding a vertex to an on-shell diagram via BCFW bridge.}
	\label{fig:bcfw-general}
\end{figure}

\paragraph{General BCFW bridging}
The BCFW bridge can be used to add a vertex to an on-shell diagram 
at a fixed number of external legs. The idea is sketched in 
Fig.~\ref{fig:bcfw-general}. 

\noindent 
The rotation matrix, as in \eqref{eq:L12rot} or \eqref{eq:L14rot}, is an element of an O$(2,\IC)$ subgroup of O$(2k,\IC)$ 
acting on the kinematic variables $\Lambda_i$. 
It is worth noting whether the rotation matrix belongs 
to the orientation preserving SO$(2k,\IC)$ subgroup of O$(2k,\IC)$ 
or the orientation reversing one. In the case depicted in 
Fig.~\ref{fig:bcfw-vertex}(a), $\s=+1$ preserves orientation. 
In contrast, in the case of Fig.~\ref{fig:bcfw-vertex}(b), 
$\s=-1$ is the one that preserves orientation.

\paragraph{Vertex as OG$_2$}
We proceed to relate the BCFW bridge to OG$_2$. 
Consider the integral,
%The OG integral for $k=2$ is %given by %\cite{Lee:2010du}
\begin{align}
\CL_{4}(\Lambda) =
\int  \frac{d^{2\times 4} C}{{\rm vol}[{\rm GL}(2)]}\frac{\delta^{3}(C\cdot  C^T)\, \delta^{4|6}(C\cdot \L)}
{(12)(23)}
\,.
\label{grass-4pt}
\end{align}
We will show that $\CL_4$ reproduces $A_4$ in 
the form \eqref{eq:bcfw-4pt-vertex-12} or \eqref{eq:bcfw-4pt-vertex-14}.
For the former, depicted in Fig.~\ref{fig:bcfw-vertex}(a), 
the choice of source legs $\{1,2\}$ naturally translates into a gauge fixing 
of the matrix $C$ with `pivot' columns $\{1,2\}$, 
\begin{align}
C = 
\begin{pmatrix}
1 & 0 & c_{13} & c_{14} \\
0 & 1 & c_{23} & c_{24} 
\end{pmatrix} \,.
\label{eq:Cfix14}
\end{align}
We can proceed in two different ways. First, as explained in \cite{Gang:2010gy}, we can solve the bosonic part of 
the kinematic delta function $\delta^{4|6}(C\cdot \Lambda)$. 
Since there are four delta functions for four variables, 
for generic values of $\l_i$ $(i=1,2,3,4)$, 
the solution to the delta function constraint is unique. Inserting the solution back to the integral, we arrive at the expression for $A_4$ in \eqref{eq:4ptint14a}. 
Alternatively, we can leave $\delta^{4|6}(C\cdot \Lambda)$ aside and 
insert the gauge-fixed $C$ \eqref{eq:Cfix14} into the orthogonality constraint delta function, 
\begin{align}
\delta^3(C\cdot C^T) =  \delta(1+c_{13}^2 + c_{14}^2)\,\delta(1+c_{23}^2 +c_{24}^2)\,\delta (c_{13}c_{23} + c_{14} c_{24}) \,,
\end{align}
which takes a similar form as \eqref{eq:P-14}. 
Taking the change of variables,
\begin{align}
\begin{pmatrix}
c_{13} & c_{14} \\
c_{23} & c_{24}
\end{pmatrix} \rightarrow
i \begin{pmatrix}
r_1 \sin{t_1} & r_1 \cos{t_1}\\
r_2 \cos{t_2} & r_2 \sin{t_2}
\end{pmatrix},
\end{align}
and integrating out $(r_1,r_2,t_2)$,  we obtain
\begin{align}
\mathcal{L}_4 = \sum_{\sigma} \int \frac{dt}{4i \sin{t}} \delta^{2|3}(\Lambda_1 + i \sin{t} \Lambda_3 + i \sigma \cos{t} \Lambda_4)\,\delta^{2|3}(\Lambda_2 + i \cos{t} \Lambda_3 - i \sigma \sin{t} \Lambda_4),
\end{align}
in agreement with \eqref{eq:bcfw-4pt-vertex-12}. 
Repeating the same analysis with pivot columns $\{1,4\}$, 
we find that
$\CL_4(\L_1,\L_2,\L_3, \L_4)$ reduces precisely to $A_4$
given by \eqref{eq:bcfw-4pt-vertex-14}.
Thus, the OG integral provides a geometric representation of the 4-point vertex. 

\subsection{Amalgamation and permutation}
\begin{figure}[htbp]
	\centering
	\includegraphics[height=3cm]{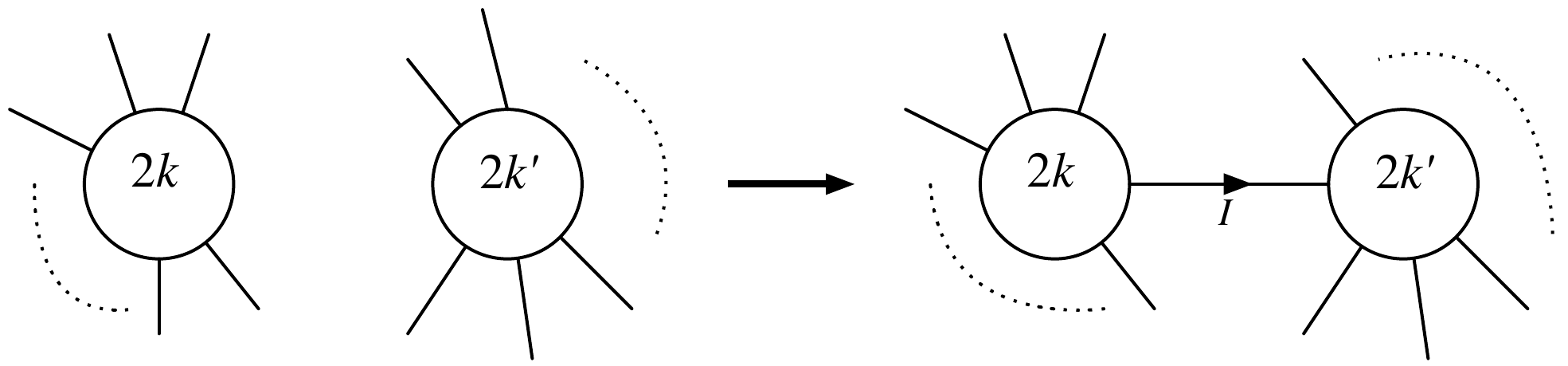}
	\caption{Gluing two on-shell diagrams.}
	\label{fig:amalgamation}
\end{figure}

An on-shell diagram for $k>2$ can be constructed by gluing two or more diagrams together. The corresponding OG$_{k > 2}$ is obtained by 
`amalgamating' a collection of $\text{OG}_2$'s by integrating out the internal lines. The amalgamation proceeds as follows (see Fig.~\ref{fig:amalgamation}).

\begin{enumerate}
	\item Put two on-shell diagrams together, preserving all external legs. The resulting $C$-matrix is a direct product of the two sub-matrices and thus lives in $\text{OG}_{k+k'}$.
	\item Pick an external line from each diagram. Identify them by setting the one as $\Lambda_I$ and the other as $-i \Lambda_I$, and perform the integral (\ref{def-internal1}). Since it reduces a number of external legs by 2, the result must be an element of $\text{OG}_{k+k'-1}$.
\end{enumerate}
Integrating out the internal line explicitly, we have
\begin{align}
	&\int d^{2|3}{\Lambda_I} \prod_{m=1}^{k}\delta^{2|3}\bigg(\sum_{i=1}^{2k-1} c_{m i} \Lambda_i + c_{m I} \Lambda_I\bigg) \prod_{n=1}^{k'} \delta^{2|3}\bigg(\sum_{j=1}^{2k'-1} c'_{n j} \Lambda'_j - i \,c'_{n I} \Lambda_I\bigg) \nonumber\\
	&\quad = c_{I1} \prod_{m=2}^{k} \delta^{2|3}\bigg(\sum_{i=1}^{2k-1} \big(c_{m i} - \frac{c_{1i}}{c_{1I}}c_{m I}\big) \Lambda_i\bigg) \prod_{n=1}^{k'} \delta^{2|3}\bigg(\sum_{j}^{2k'-1} c'_{n j} \Lambda'_j + i \sum_{i=1}^{2k-1} \frac{c_{1i}}{c_{1I}} c'_{n I} \Lambda_i\bigg)\,.
\end{align}
The resulting matrix $\widetilde{C}$ is represented by %\footnotemark[1]
\begin{equation}
	\arraycolsep=10pt
	\def\arraystretch{2}
	\widetilde{C}_{\tilde{m} \tilde{\imath}} = \left(\begin{array}{>{\displaystyle}c|>{\displaystyle}c}
	c_{m i} - \frac{c_{m I}}{c_{1I}} c_{1i} & 0 \\[0.6em] \hline
	i\,\frac{c'_{n I}}{c_{1I}} c_{1i} & c'_{n j}
	\end{array}\right)_{\tilde{m} \tilde{\imath}} \,,
	\label{eq:amalgamation}
\end{equation}
where $\tilde{m}$ runs over \{$m = 2 \cdots k$, $n = 1 \cdots k'$\} and $\tilde{\imath}$ over \{$i = 1 \cdots 2k-1$, $j = 1 \cdots 2k'-1$\}. 
As shown in \cite{Huang:2013owa}, it is straightforward to verify 
that $\widetilde{C}$ respects the orthogonality condition $\widetilde{C} \cdot \tilde{C}^T = 0$ if $C$ and $C'$ satisfy the same condition,
\begin{equation}
	\sum_{i=1}^{2k-1}  c_{m i}c_{n i} +  c_{m I} c_{n I} = 0 \,,\hspace{0.5cm}
	\sum_{j=1}^{2k'-1} c'_{m j}c'_{n j} + c'_{m I} c'_{n I} = 0\,.
\end{equation}

\begin{figure}[htbp]
	\centering
	\includegraphics[height=3.2cm]{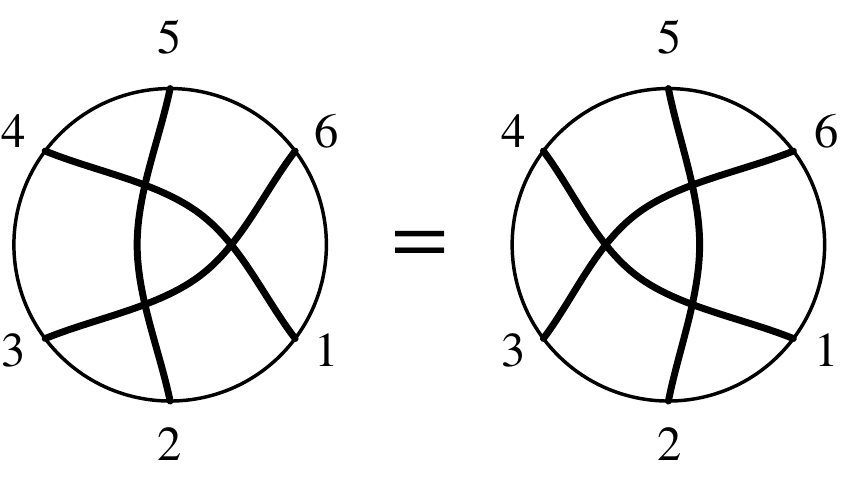}
	\caption{Yang-Baxter equivalence move}
	\label{fig:yangbaxter1}
\end{figure}

\paragraph{Equivalences move and reducible diagrams} 
Different on-shell diagrams corresponding to the same amplitude 
can be related to each other through a series of equivalence moves. 
The elementary move for ABJM amplitudes is 
the Yang-Baxter-like move depicted in Fig.~\ref{fig:yangbaxter1}. 
The lines 1 to 6 in the figure may be internal or external. 
In view of the BCFW bridging, the equivalence relation simply amounts 
to two different Euler angle decomposition of the same 
SO$(3)$ rotation matrix. 

\begin{figure}[htbp]
	\centering
	\includegraphics[height=2.2cm]{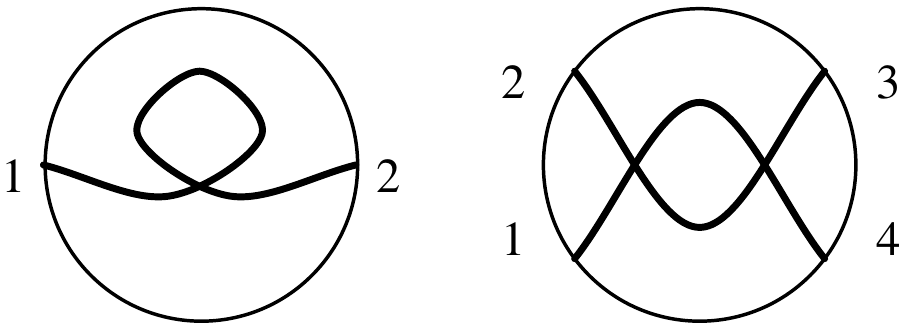}
	\caption{Diagrams containing bubbles}
	\label{fig:bubble}
\end{figure}

As the name `Yang-Baxter' suggests, the equivalence 
move leaves the permutation among external particles invariant. 
If we draw all possible diagrams with the same permutation content, 
we may encounter bubble diagrams such as those in Fig.~\ref{fig:bubble}. 
It was shown in \cite{Huang:2013owa} that 
the bubbles can be completely factorized from the rest of the on-shell diagram, leaving an integral with a $d\log$ measure. 
In the rest of this paper, we will work exclusively with bubble-free diagrams. 

%\newpage
\section{Reality and Positivity of Orthogonal Grassmannian \label{sec:OG}}
 
The orthogonal Grassmannian $\text{OG}_k$ is a subspace of $\text{G}(k,2k)$
restricted by the orthogonality constraint. 
In this section, we shall examine the notion of reality and positivity for $\text{OG}_k$.
The real slice of $\text{OG}_k$ is determined by the reality condition on external kinematic variables.
For discussion on positivity, a particular reality condition, 
called the `split signature' \cite{Huang:2013owa} condition, 
turns out to be the most convenient. 
Performing a `Wick rotation' on the kinematic and the BCFW variables, we can rewrite the BCFW bridging rule 
with manifestly real and positive coordinates on OG$_k$. 
Following \cite{ArkaniHamed:2012nw,Huang:2013owa}, 
we spell out the conversion rule 
which enables us to read off the postive $C$-matrix directly from an on-shell diagram without going through BCFW bridging one at a time. 
%The results in this section lay the foundation for the positive stratifications of POG to be discussed in the next section.

\subsection{Complex OG}

Here we review the geometry of OG$_k$ 
on which the integral \eqref{grass0} is defined \cite{Gang:2010gy}. 
Recall that the ordinary Grassmannian G$(k,n)$ is the moduli space 
of $k$-planes in $n$ dimensions. In the standard matrix representation, 
G$(k,n)$ is described by a $(k \times n)$ matrix $C$ with rank $k$ 
subject to the `gauge symmetry' $C \sim g \,C$ with $g \in {\rm GL}(k)$.
OG$_k$ is a subspace of G$(k,2k)$ 
subject to an `orthogonality' constraint 
$(C\cdot C^T)_{mn} = C_{m i} C_{n i} = 0$ 
$(m=1,\ldots,k;i=1,\ldots,2k)$. 
The constraint and the GL$(k)$ gauge symmetry determine the dimension of OG$_k$ as
\begin{align}
\dim_\IC\left[{\rm OG}_k\right] = 2k^2 - k^2 - \frac{k(k+1)}{2} = \frac{k(k-1)}{2} \,.
\end{align}
It is also known that OG$_k$ is isomorphic to the coset, %\cite{Gang:2010gy},  
%\begin{align}
${\rm OG}_k = {\rm O}(2k)/{\rm U}(k)$. 
%\end{align}
Since O$(2k)$ contains two disjoint SO$(2k)$ components, 
OG$_k$ is also decomposed into two disjoint subspaces. 
We will call them two `branches' of OG$_k$. 
%and introduce the short-hand notation OG$_k^{\pm}$. 

Let us discuss how the two branches are defined in terms of coordinates.
Recall that the Pl\"ucker coordinates of G$(k,n)$ 
are determinants of $(k\times k)$ submatrices of $C$, 
regarded as homogeneous coordinates of some projective space. 
%The ratios among the determinants serve as GL$(k)$ invariant coordinates. 
The Pl\"ucker coordinates are subject to quadratic algebraic relations 
originating from linear dependencies among the columns of $C$.  
Coming back to OG$_k$, 
let $I = \{ i_1, i_2, \ldots, i_k \}$ 
be an ordered set of indices labeling $k$ distinct columns of $C$, 
and $M_I = (i_1 , i_2 , \cdots, i_k) := \det( C_{i_1}, C_{i_2}, \cdots, C_{i_k})$ be the corresponding 
Pl\"ucker coordinate. 
As noted in \cite{Lee:2010du}, the orthogonality constraint $C\cdot C^T=0$ 
imposes {\em linear} relations among $M_I$'s in addition to 
the quadratic relations for G$(k,n)$. 
Define the complement of $I$ by $\bar{I}= \{\bar{\imath}_1,\bar{\imath}_2,\ldots, \bar{\imath}_k\}$ such that $\{i_1,\ldots, i_k,\bar{\imath}_1,\ldots, \bar{\imath}_k\}$ is an {\em even} permutation of 
$\{1,2, \ldots, 2k\}$. 
The linear relation can be written as
\begin{align}
\label{eq:branch}
M_I = \s (i^k) M_{\bar{I}} \,.
\end{align}
The overall sign factor $\s=\pm 1$ defines the two branches of OG$_k$. 
This linear relation implies a quadratic relation for consecutive minors 
on both branches \cite{Lee:2010du}, 
\begin{align}
M_i M_{i+1} = M_{i+k} M_{i+1+k} (-1)^{k-1} \,.
\label{quad-consec}
\end{align}
%Note that 
The orthogonality constraint can 
be expressed in terms of Pl\"ucker coordinates as
\begin{align}
\sum_a (i_1,\cdots,i_{k-1},a)(j_1,\cdots,j_{k-1},a) = 0\,,
\label{ortho-plu}
\end{align}
where $a$ runs over all columns of the $C$-matrix. 

For $k=2$, we can solve all the relations explicitly. 
The two branches are defined by
\begin{align}
\mbox{OG}_k^{\pm}\; : \quad (M_{12}, M_{23}, M_{31}) = \mp (M_{34},M_{14},M_{24}) \,. 
\end{align}
Combining this with the Schouten identity, 
\begin{align}
M_{12} M_{34} + M_{23} M_{14} + M_{31} M_{24} = 0 \,, 
\end{align} 
and renaming the coordinates as $(X,Y,Z) = (M_{14}, M_{24}, M_{34})$, 
we find that each of the two branches, OG$_2^\pm$, is described by 
an algebraic variety,
\begin{align}
\{ (X,Y,Z) \in \IC\IP^{2} \,|\, X^2+ Y^2 + Z^2 = 0 \} \,, 
\label{OG2-alg}
\end{align}
which is topologically a $\IC\IP^1$. We can compare this with 
the coset description, 
\begin{align}
\mbox{OG}_2^+ = SO(4)/U(2) \simeq SU(2)\times SU(2) / U(1)\times SU(2) \simeq 
SU(2)/U(1)  = \IC\IP^1 \,.
\end{align}
Repeating the algebraic analysis for higher $k$ would be possible but quite cumbersome. 
For $k=3$, we can use the coset description to find 
\begin{align}
\mbox{OG}_3^+ = SO(6)/U(3) \simeq SU(4)/U(3) = \IC\IP^3 \,.
\end{align}
%{\bf (Resolve the $\IZ_2$ subtlety in ${\rm SO}(2k) = {\rm Spin}(2k)/\IZ_2$.)}

\subsection{Reality and positivity of OG}
\label{sub:POG}

\paragraph{Reality}

In the spinor helicity formulation $p^{\a\b} = \l^\a \l^\b$ with real momentum, the spinor $\l^\a$ should be real or purely imaginary. 
Our convention is such that the spinor $\l^\a$ is real for outgoing particles and purely imaginary for incoming particles. If we want to work with strictly real momenta, we have to assign reality conditions on each of the external legs. 

Momentum conservation forces all on-shell diagrams to have the same number of incoming and outgoing particles. 
First, the elementary 4-vertex (\ref{def-vertex1}) does not vanish only if 
two of the particles are incoming and the other two are outgoing. 
If all four particles are outgoing, the total momentum, $\lambda_1 \lambda_1 + \lambda_2 \lambda_2 + \lambda_3 \lambda_3 + \lambda_4 \lambda_4$ is positive definite or negative definite, respectively, so cannot vanish. 
If particle 1, 2, 3 are outgoing and 4 incoming, $\lambda_1 \lambda_1 + \lambda_2 \lambda_2 + \lambda_3 \lambda_3$ generically has rank two
while $\lambda_4 \lambda_4$ has rank one, in conflict with momentum conservation. The other two unbalanced cases (4 incoming or 3 incoming + 1 outgoing) can be treated similarly.
Next, an internal line \eqref{def-internal1} always connects an incoming particle and an outgoing particle, hence the balance between incoming and outgoing particles continue to hold for arbitrary on-shell diagrams.

The reality conditions for the kinematic variables $\lambda_i$ naturally translate into the reality conditions for the matrix $C$ representing a point on OG$_k$. The linear delta function $\delta(C\cdot \Lambda)$ 
in \eqref{grass0} requires that  
\begin{align}
\sum_{i=1}^{2k} C_{mi} \lambda_i = 0 \,.
\end{align}
We may use the GL$(k)$ gauge symmetry to make the $k$ columns 
corresponding to incoming particles to form a $(k\times k)$ identity matrix, 
and denote the other $k$ columns by a non-trivial $(k\times k)$ matrix: 
\begin{align}
\lambda_i + c_{i\bar{\jmath}} \lambda_{\bar{\jmath}} = 0 \,, 
\label{real-gauge}
\end{align}
where the index $i$ runs over incoming particles and $\bar{\jmath}$ over outgoing particles. Reality of $\lambda$ implies that $i c_{i\bar{\jmath}}$ is a real matrix. The orthogonality constraint $C\cdot C^T=0$ then implies 
that $ic_{i\bar{\jmath}}$ is an element of O$(k,\IR)$. Thus, the reality condition for $\lambda_i$ defines a real slice of OG$_k$. The two 
disconnected components of O$(k,\IR)$ correspond 
to the two branches of OG$_k$. 
For instance, the real slices of $k=3$ are two copies of SO$(3)= \IR \IP^3$. 

\begin{figure}[htbp]
	\centering
	\includegraphics[height=4.5 cm]{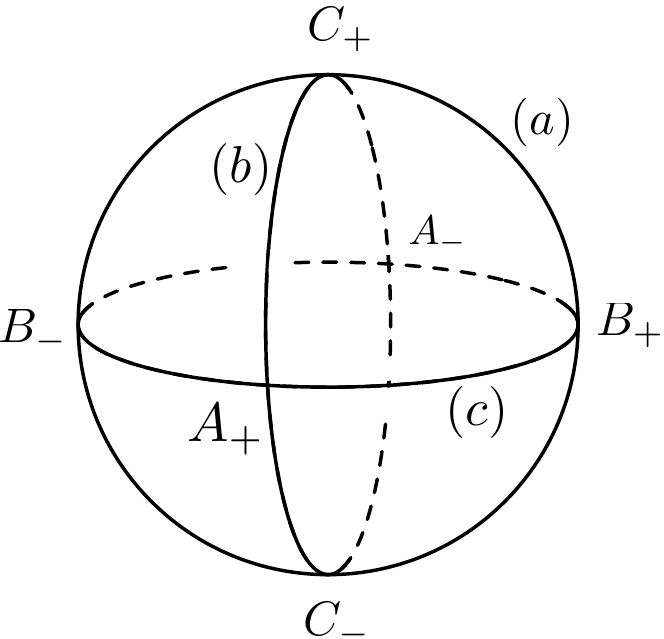}
	\caption{Real slices of a OG$_{2+}$.}
	\label{fig:real-k2}
\end{figure}

We find it instructive to give a gauge invariant, geometric description of 
the reality conditions. 
We will focus on the simplest example for $k=2$. In the algebraic description \eqref{OG2-alg}, in a coordinate patch with $Z \neq 0$, 
we can consider four distinct reality conditions: 
\begin{align}
\begin{array}{c|cccc}
& (a) & (b) & (c) & (d) \\ \hline
X/Z & i\IR & \IR & i\IR & \IR \\
Y/Z & \IR & i\IR & i\IR & \IR 
\end{array}
\end{align}
But, the condition (d) with $X^2+Y^2+Z^2 = 0$ yields an empty set in $\IC\IP^2$. 
The three remaining real slices are identified with 
great circles, $S^1 = {\rm SO}(2,\IR)$, embedded in $S^2 = \IC\IP^1$ as depicted in Fig.~\ref{fig:real-k2}. 
The intersections among different real slices 
are denoted as
\begin{align}
A_{\pm}\, :\, X = 0,\;\; Y/Z = \pm i \,, 
\;\;
B_{\pm}\, : \,  Y= 0,\;\; Z/X = \pm i \,, 
\;\; 
C_{\pm}\, : \,   Z= 0,\;\; X/Y = \pm i \,.
\end{align}
We should emphasize that the reality condition 
is a gauge invariant notion. 
Although it is sometimes useful to align the gauge choice 
with the reality condition as in \eqref{real-gauge}, 
other gauge choices might be more convenient for certain purposes. 
In Sec.~\ref{sec:positroid}, we will use gauge choices different from \eqref{real-gauge}.

\paragraph{Positivity}

As pointed out in \cite{Huang:2013owa}, the `split signature' reality condition, 
in which all odd-labelled particles are incoming and all even-labelled 
ones outgoing, deserves a special attention. It is the only reality condition 
that respects the cyclic symmetry of $A_{2k}$ \eqref{A-cyclic}. 
Moreover, the split signature allows for a simple notion of `positivity'. 
Following \cite{Huang:2013owa}, we do a `Wick rotation' on the $C$-matrix and $\lambda_{2i-1}$ simultaneously such that all elements of $C$ and all $\lambda_i$ are real, while the orthogonality constraint takes the form, 
\begin{align}
C\cdot \eta \cdot C^T = 0 \,,
\quad 
\eta_{ij} = (-1)^i \delta_{ij} = {\rm diag}(-,+,\cdots,-,+) \,.
\end{align}
In this convention, the positivity defined in \cite{Huang:2013owa} 
asserts that all ordered minors of $C$ are non-negative.
\footnote{A similar notion for G$(k,n)$ is called `totally non-negative' 
in mathematics literature \cite{Postnikov:2006kva,Knutson:2011,Lam:2007}. For brevity, we will write `positive' 
in place of `totally non-negative'.}
This particular definition of positivity picks out 
one of the two branches of OG$_k$ at each $k$. For instance, 
we have
\begin{align}
(1,2,\ldots,k)=(k+1,k+2,\ldots,2k)
\quad 
\mbox{and}
\quad
(1,3,\ldots,2k-1) = (2,4,\ldots,2k) \,.
\label{positive-branch}
\end{align}
It was shown in \cite{Huang:2013owa} that the definition of positivity 
is compatible with all essential properties of OG such as
\eqref{quad-consec} and \eqref{ortho-plu}.

These results strongly suggest that 
the positive orthogonal Grassmannian (POG) 
has the same dimension as the real slice of OG. 
%, isomorphic to SO$(k,\IR)$. 
One of the main goals of this paper 
is to introduce a complete set of coordinate patches for POG for all $k$. 
We give a simplest example ($k=2$) here for illustration 
and discuss the general construction in Sec.~\ref{sec:positroid}. 
Let us choose a gauge such that 
\be
C = 
\begin{pmatrix}
1 & 0 & c_{13} & c_{14} \\
0 & 1 & c_{23} & c_{24} 
\end{pmatrix} \,.
\ee
In the split signature, $\eta = {\rm diag}(-,+,-,+)$, the orthogonality constraint $C\cdot \eta \cdot C^T = 0$ gives
\begin{align}
-1 - c_{13}^2 + c_{14}^2 = 0 \,, 
\quad 
1 - c_{23}^2 + c_{24}^2 = 0 \,,
\quad 
-c_{13} c_{23} + c_{14} c_{24} = 0 \,.
\end{align}
Positivity requires that 
\begin{align}
c_{13}, c_{14} \le 0 \,,
\quad 
c_{13}, c_{14} \ge 0 \,.
\end{align}
The complete solution to this problem is 
\begin{align}
\begin{pmatrix}
c_{13} & c_{14} \\
c_{23} & c_{24} 
\end{pmatrix} 
= 
\begin{pmatrix}
-\sinh t & -\cosh t \\
\cosh t & \sinh t 
\end{pmatrix} \,,
\qquad t \ge 0 \,.
\end{align}
As $t$ approaches $\infty$, we can take a gauge transformation 
to find
\begin{align}
C = \begin{pmatrix}
1 & -\sinh t  \\
0 & \cosh t  
\end{pmatrix}
\begin{pmatrix}
1 & \tanh t &  0 & -{\rm sech} t \\
0 & {\rm sech} t & 1 & \tanh t \\
\end{pmatrix} 
\rightarrow 
\begin{pmatrix}
1 & 1 &  0 & 0 \\
0 & 0 & 1 & 1 \\
\end{pmatrix} \,.
\label{k2-inf}
\end{align}
Including the `point' at $t=+\infty$, 
the full geometry of POG$_2$ is an interval with two end-points included. 
In Fig.~\ref{fig:real-k2}, the POG$_2$ is identified with the interval $\overline{A_- C_+}$. 

%\newpage

\subsection{Conversion rule}
\label{sec:conversion}

Recall that the elementary vertex, $A_4$, is resolved in two ways, \eqref{eq:bcfw-4pt-vertex-12} and \eqref{eq:bcfw-4pt-vertex-14}, according to the choice of pivot columns for OG$_2$. The first one \eqref{eq:bcfw-4pt-vertex-12} 
contains a delta-function, 
\begin{align}
%(12\rightarrow 34) \; : \; 
\delta^{2|3}(\L_1 + i\sin t \L_3 + i\s \cos t \L_4) \,
\delta^{2|3}(\L_2 + i\cos t \L_3 - i\s \sin t \L_4) \,.
\end{align}
We perform a Wick rotation on the odd-labelled particles as 
$\Lambda_{2i-1} \rightarrow -i \Lambda_{2i-1}$ 
and on the BCFW variables as $t \rightarrow i t$. 
After the Wick rotation, up to an overall phase, the delta-function becomes
\begin{align}
%(12\rightarrow 34) \; : \; 
\delta^{2|3}(\L_1  - \sinh t \L_3 - \s \cosh t \L_4) \,
\delta^{2|3}(\L_2 + \cosh t \L_3 + \s \sinh t \L_4) \,.
\label{vfac1}
\end{align}
Similarly, in the other case \eqref{eq:bcfw-4pt-vertex-14}, the 
Wick rotation gives 
\begin{align}
&\delta^{2|3}(\L_1 + i \s \cos t \L_2 - i\sin t \L_3) \,
\delta^{2|3}(\L_4 - i\s\sin t \L_2 - i \cos t \L_3) \,.
\nn \\
\rightarrow \;\; &\delta^{2|3}(\L_1  - \s \cosh t \L_2 + \sinh t \L_3) \,
\delta^{2|3}(\L_4 +\s\sinh t \L_2 - \cosh t \L_3) \,.
\label{vfac2}
\end{align}
We apply the same Wick rotation to the internal line \eqref{def-internal1} as well, 
\begin{align}
\delta(i\Lambda_1 + \Lambda_2) 
\quad \rightarrow \quad 
\delta(\Lambda_1 + \Lambda_2) \,,
\label{internal-Wick}
\end{align}
where $\L_1$ and $\L_2$ represent odd and even multiplets, respectively. 

\begin{figure}[htbp]
	\centering
	\includegraphics[height=2.6cm]{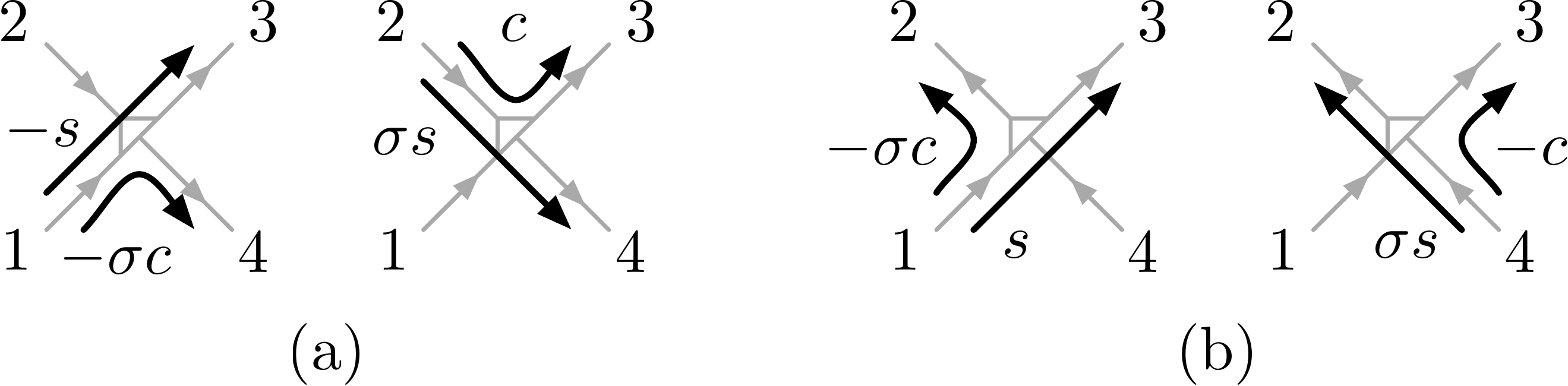}
	% \caption{Diagrammatic Vertex Rules for $(-f_{v})$ under the Kinematic Gauge}
	\caption{Vertex factors $f_{v}$ in the canonical gauge 
	to be used in the conversion rule.}
	\label{fig:sc-rule}
\end{figure}

As explained in \cite{ArkaniHamed:2012nw,Huang:2013owa}, 
we can collect the linear relations imposed by the delta-functions 
and read off the components of the $C$-matrix 
without going through BCFW bridging and amalgamation one at a time. 
To find the component $c_{i\bar{\jmath}}$ in the gauge-fixed form,
%\begin{align}
$\l_i + c_{i\bar{\jmath}} \l_{\bar{\jmath}} = 0$,
%\end{align}
we trace all possible paths $p$ from the source $i$ to the sink $\bar{\jmath}$ in the corresponding on-shell diagram. 
Each internal line \eqref{internal-Wick} contributes a factor of $(-1)$, since $\delta(\l_1+ \l_2)$ implies $\l_1 = -\l_2$.
At each vertex, we pick up $(-f_v)$, where 
$f_v$ is one of the matrix elements of the $(2\times 2)$ matrix defined at the vertex, $\l_i + c^{(v)}_{i\bar{\jmath}} \l_{\bar{\jmath}}=0$, chosen by how the path traverses the vertex. 
\footnote{The minus sign in $(-f_v)$ arises because we view 
the relation $\l_i + c^{(v)}_{i\bar{\jmath}} \l_{\bar{\jmath}}=0$ 
as $\l_i = - c^{(v)}_{i\bar{\jmath}} \l_{\bar{\jmath}}$. 
The overall minus sign right after the first equality sign in \eqref{eq:diagramrule} is inserted for the same reason.}
The vertex factors in the two cases of BCFW bridging considered above 
are summarized in Fig.~\ref{fig:sc-rule}. 
The factors in Fig.~\ref{fig:sc-rule}(a) are read off from \eqref{vfac1} and those in Fig.~\ref{fig:sc-rule}(b) from \eqref{vfac2}.
Combining the contributions from internal lines and vertices, 
we arrive at a simple conversion rule for $c_{i\bar{\jmath}}$:
\begin{equation}
	c_{i\bar{\jmath}} = - \sum_{p \in \{i\rightarrow \bar{\jmath}\}} \left(\prod_{l \in p} (-1) \cdot \prod_{v \in p} (-f_{v})\right) = \sum_{p \in \{i\rightarrow \bar{\jmath}\}}  \left(\prod_{v \in p} f_{v}\right)\,.
	\label{eq:diagramrule}
\end{equation}
The matrix elements are manifestly real. The factors of $(-1)$ 
have cancelled out completely, since a path always traverse 
$n_p$ internal lines and $n_p+1$ vertices. We will 
study how positivity restricts the matrix elements in the next section. 

%\newpage

\section{Positroid Stratification}
\label{sec:positroid}

The `positroid stratification' of G$_+(k,n)$ was developed in \cite{Postnikov:2006kva,Knutson:2011} and brought to physics in \cite{ArkaniHamed:2012nw}. 
It relates the combinatorics of 
on-shell diagrams to the geometry of G$_+(k,n)$. At an intermediate step, 
a notation similar to Young tableaux plays a crucial role, which encodes the linear dependency among the columns of the matrix representative of G$_+(k,n)$. 
In this section, we shall develop a similar story for the positive stratification of OG. 
Some partial results in this direction were obtained in \cite{Huang:2013owa}.

As a first step, we introduce an auxilary aid called `OG tableau' which encodes the combinatorics of on-shell diagrams.
It naturally provides a set of canonical gauge choices for the $C$-matrix 
such that the restriction imposed by positivity takes a simple form.
With the help of the OG tabeaux, we construct the `canonical' coordinate system, which exhibits positivity by construction 
for each cell of $\text{POG}_k$ for all $k$. 
We verify that the canonical coordiates obtained from 
the positive stratification agrees with the ones 
given by the conversion rule derived in Sec.~\ref{sec:OG}.  

In the last subsection, we turn to the mathematics of POG$_k$. 
It is known that the positive Grassmannian G$_+(k,n)$ forms a combinatorial polytope called `Eulerian poset' for each $(k,n)$ \cite{Williams:2007}. 
The graded counting of OG tableaux suggests that POG$_k$ may also define an Eulerian poset for each $k$. Geometrically, it seems plausible that POG$_k$ has a topology of a ball. We verify 
this conjecture for $k=2,3$. Finally, we give a preliminary discussion on the boundary operation on POG$_k$. 
We expect that a more complete study of the boundary operation 
will help us better understand the topology and geometry of POG.

\subsection{OG tableaux}
\label{sub:cell-decomposition}

\begin{figure}[h]
  \centering
  \includegraphics[height=2.5cm]{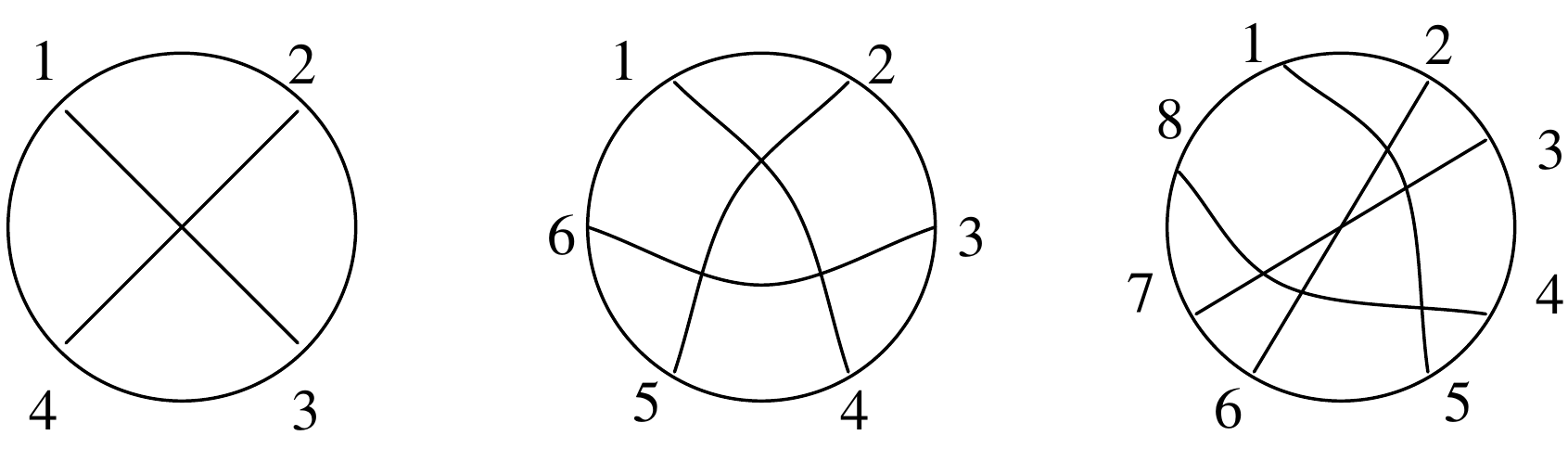}
  \caption{Top-cell diagrams for $k=2, 3, 4$}
  \label{fig:top-cell}
\end{figure}

As we discussed earlier, on-shell diagrams are determined by a splitting of $\{1,2,\ldots, 2k\}$ into $k$ pairs of integers, $\{(a_1 b_1), \cdots, (a_k b_k)\}$. 
By convention, we set $a_m < b_m$ for all $m$. Barring bubbles and modulo Yang-Baxter equivalence moves, the diagrams are in one-to-one correspondence with 
the pairings. The total number of inequivalent diagrams are 
$(2k)!/( 2^k k!) = 1,3,15,105, \cdots$. Some subclasses of diagrams 
are easy to enumerate. 
For a given $k$, there is a unique `top' diagram with the maximal number, $k(k-1)/2$, of vertices (see Fig.~\ref{fig:top-cell}). All external legs in a top diagram are paired diagonally. 

\begin{figure}[h]
  \centering
  \includegraphics[height=4.5cm]{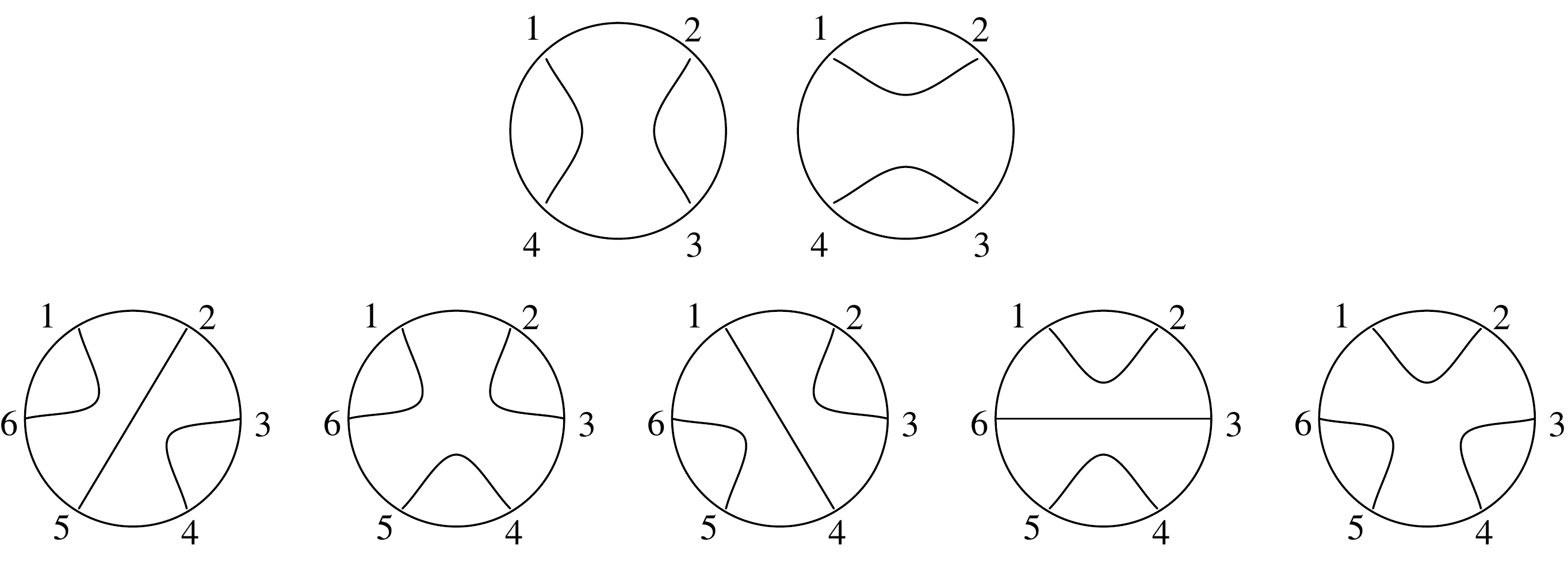}
  \caption{Bottom-cell diagrams for $k=2$ (upstairs) and $k=3$ (downstairs)}
  \label{fig:bottom-cell}
\end{figure}

\noindent
At the opposite extreme, there are `bottom' diagrams with no vertex (see Fig.~\ref{fig:bottom-cell}). The counting of non-intersecting diagrams connecting $2k$ cyclically ordered points is an elementary problem in combinatorics. The answer is the $k$'th Catalan number ${\rm C}_k$ \cite{catalan},
\begin{align}
{\rm C}_{k} 
= \frac{1}{k+1} \binom{2k}{k}
= \frac{(2k)!}{(k+1)!k!}  = 1, 2, 5, 14, 42 , \cdots  \,.
\end{align}
To enumerate the diagrams with intermediate number of vertices, 
and relate them to subspaces of OG$_k$, 
we introduce a new notation called `OG tableaux'. 
There are two related versions of the tableaux: `unfolded' and `folded'.

\paragraph{Unfolded tableaux}

\begin{figure}[htbp]
  \centering
  \includegraphics[height=4cm]{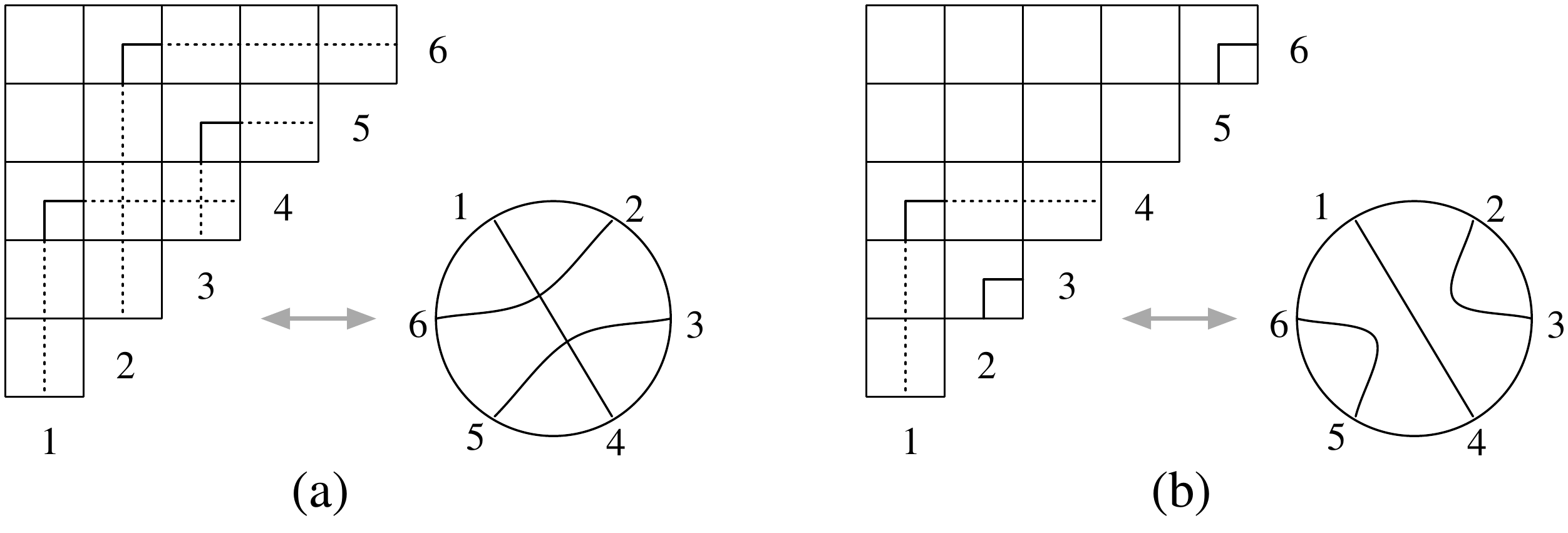}
  \caption{Examples of unfolded $\text{OG}_3$ tableaux}
  \label{fig:tableaux-unfolded}
\end{figure}

Fig.~\ref{fig:tableaux-unfolded} illustrates how to map an on-shell diagram to a tableau with an example. We first prepare the off-diagonal upper-left half of a ($2k\times 2k$) chessboard. The diagonal boxes of the chessboard are numbered from 1 to $2k$, with 1 placed at the lower-left corner and $2k$ at the upper-right corner. The empty tableau contains $2k(2k-1)/2$ boxes, in one-to-one correspondence with a pair chosen from $\{1,\ldots,2k\}$. 
For each pair $(a_m b_m) \in \{(a_1 b_1), \cdots, (a_k b_k)\}$, 
we put a `hook' on the corresponding box. 
If we extend the right/lower arm of the hook horizontally/vertically toward the diagonal, we recover precisely a copy of the on-shell diagram. 
So far, the only novelty of the tableau notation is that 
it defines a canonical way to fix the Yang-Baxter ambiguity. 

\begin{figure}[htbp]
  \centering
  \includegraphics[height=4cm]{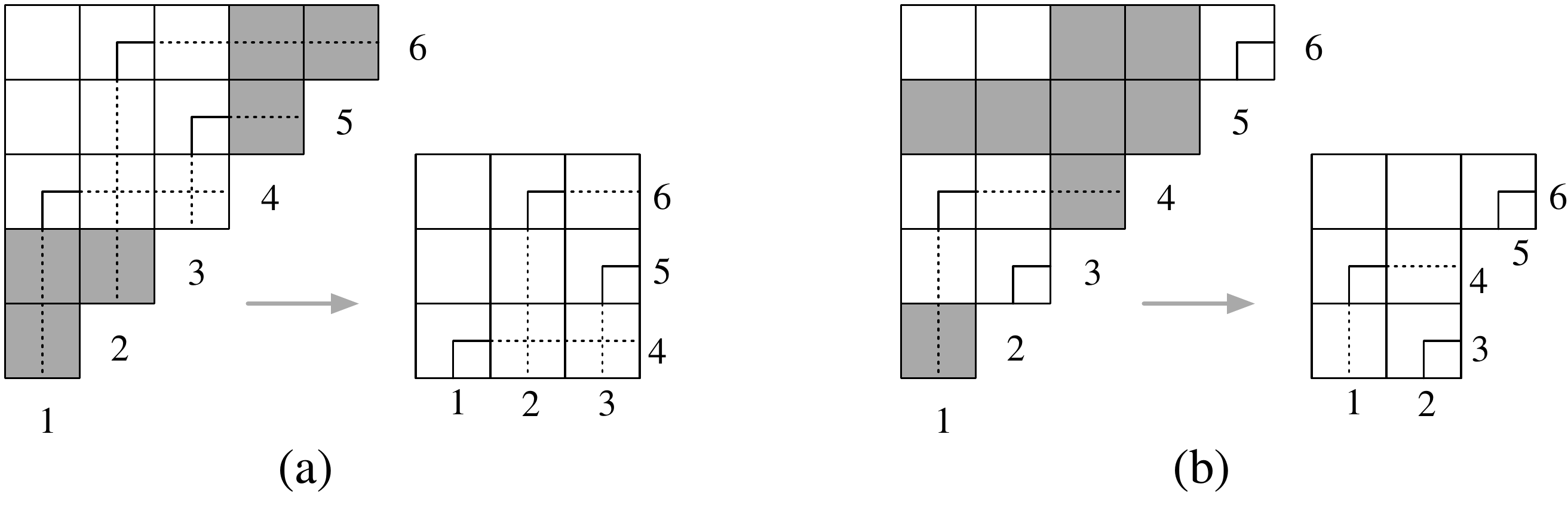}
  \caption{Folding the $\text{OG}_3$ tableaux}
  \label{fig:tableaux-folding}
\end{figure}

\paragraph{Folded tableaux} It is possible to `fold' the unfolded tableaux 
without reducing its information content. We begin with examining each of the $(k-1)$ columns and $(k-1)$ rows of the unfolded tableaux. If a column/row contains no hook, all the boxes in the column/row are removed. 
The surviving boxes can be moved in horizontal or vertical directions 
and fit into a ($k\times k$) chessboard.  
See Fig.~\ref{fig:tableaux-folding} for an illustration. 

\noindent
The content of an on-shell diagram is preserved through 
the folding procedure, so the map between on-shell diagrams and folded tableaux is still bijective.
The labels for source legs $\{a_i\}$ are attached to the boundaries at the bottom edges 
of the folded tableau, while those for sink legs  $\{b_i\}$ are attached to the right edges.  
Thus, on-shell diagrams with the same set of sink/source legs 
share the same configuration of boxes for the folded tableaux, 
but are distinguished by the placement of hooks. 
 
\begin{figure}[htbp]
	\centering
	\includegraphics[width=15cm]{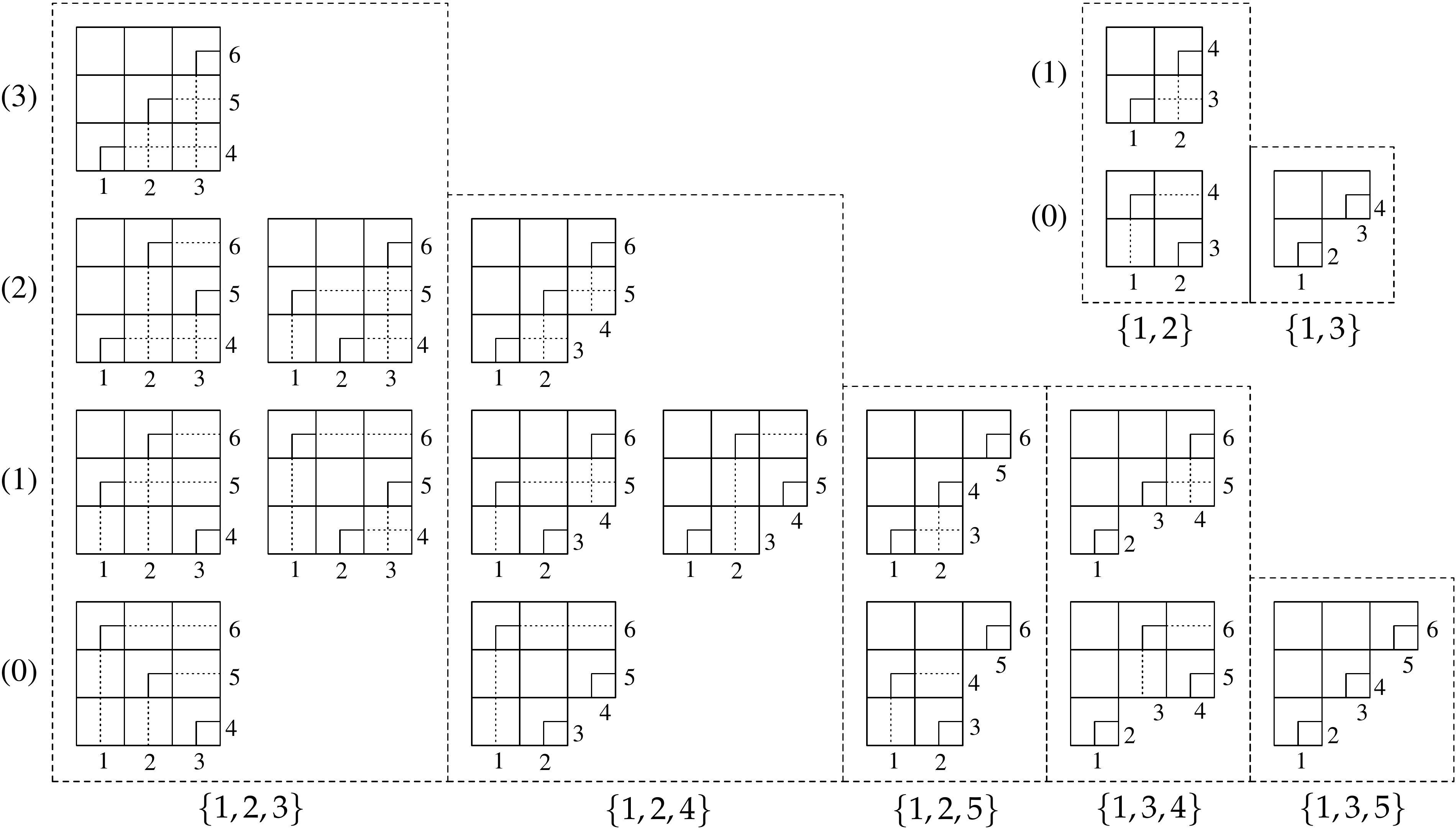}
	\caption{All diagrams for $k=2,3$ classified by folded tableaux and levels}
	\label{fig:tableaux-all-23}
\end{figure}

One of the fundamental feature of an on-shell diagram is its number of vertices (`level'). 
We can classify the on-shell diagrams according to the type 
of the corresponding (folded) tableau and the level. 
The full classification for $k=2,3$ is given in Fig.~\ref{fig:tableaux-all-23}.

\begin{figure}[htbp]
  \centering
  \includegraphics[height=2.5cm]{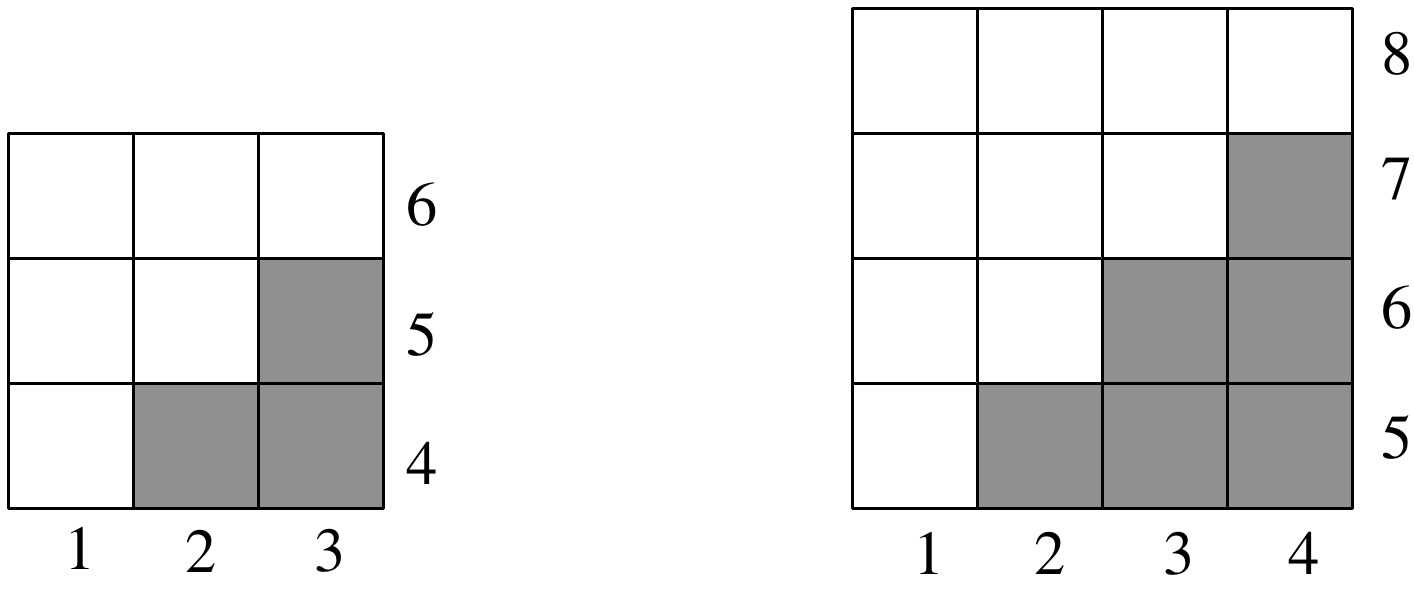}
  \caption{Shaded boxes are removable.}
  \label{fig:tableaux_rule}
\end{figure}

We introduced the unfolded tableaux first and 
switched to the folded tableaux for a pedagogical reason. But, it is no more  difficult 
to work directly with the folded tableaux. We shall construct the $\text{OG}_k$ tableaux as follows (see Fig.~\ref{fig:tableaux_rule}).
\begin{enumerate}
  \item Draw a $(k \times k)$ chessboard. 
  \item Remove some boxes among the $k(k-1)/2$ boxes in the lower-right off-diagonal half, such that the remaining boxes form a Young tableau of at least $k(k+1)/2$ boxes. 
  \item Assign $1$ to $2k$ to the bottom and right edges of the stack of boxes in order.
  \item Mark a box with a hook for each column, under the restriction that each row should contain one and only one hook.
\end{enumerate}
Each marked tableau gives an on-shell diagram. Its level can be determined as follows.
\begin{enumerate}
  \item Count how many boxes were removed from the $(k \times k)$ chessboard.
  \item Compute the number of row permutations required to arrange marked boxes diagonally from the bottom left corner to the top right.
  \item Add up the two numbers, then subtract it from $k(k-1)/2$.
\end{enumerate}
As a special case, a top diagram requires no removal of boxes 
or rearrangement of marked boxes. So, the prescription above 
gives the expected level $k(k-1)/2$.
The readers are invited to test the prescription against less trivial 
examples in Fig.~\ref{fig:tableaux-all-23}.

%\newpage

\subsection{Canonically positive coordinates}
\label{sub:positive-coord}

Positroid stratification relates the combinatorics of 
on-shell diagrams and OG tableaux to the geometry of POG$_k$. 
Each tableau is mapped to a subspace of POG$_k$. 
Importing the terminology 
from the positroid stratification of G$(k,n)$ \cite{Postnikov:2006kva,Knutson:2011}, we will call the subspaces `cells' of POG$_k$. 
The number of vertices of an on-shell diagram (level of its tableau) equals the dimension of the cell. In this subsection, we will introduce a canonical 
coordinate system to the cells. The cells sharing the same  
unmarked tableau (see Fig.~\ref{fig:tableaux-all-23}) will share a common coordinate patch. 

\paragraph{From tableaux to matrices}

\begin{figure}[htbp]
  \centering
  \includegraphics[height=1.7cm]{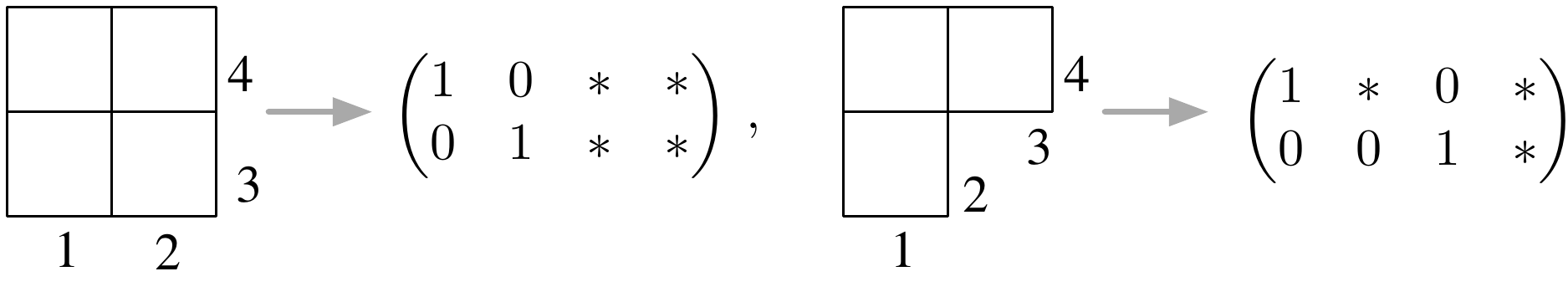}
  \caption{Source legs of tableaux translate into pivot columns of $C$-matrices}
  \label{fig:row-echelon-2}
\end{figure}

The (folded) OG tableaux reveals a decomposition of OG$_k$ similar 
to the standard Schubert decomposition of G$(k,n)$. 
The source legs translate into `pivot' columns. 
Let $\{p_m\}$ $(m=1,\ldots,k)$ be a monotonically increasing labels for the source legs. We put $C$ into a row-echelon form by setting $C_{m,p_m} = 1$, $C_{m,i < p_m}=0$ and $C_{n\neq m,p_m}=0$. An example is given in Fig.~\ref{fig:row-echelon-2}. 

\noindent
Note that using the GL$(k)$ gauge symmetry, we can write every element of $\text{OG}_k$ in the row-echelon form.
The whole $\text{OG}_k$ can be written as a disjoint union, $\text{OG}_k = \bigsqcup_{\lambda} \Omega_\lambda$, where $\lambda$ runs over unmarked tableaux. One can interpret $\Omega_\lambda$ as disjoint coordinate patches, covering the whole $\text{OG}_k$, then an on-shell diagram always belongs to a particular $\lambda$.

Unlike the Schubert decomposition of G$(k,n)$, 
in the current setup, 
the orthogonality constraint restricts the allowed set of pivot columns. 
Since the folded tableaux descend from the unfolded tableaux which 
in turn are copied from allowed on-shell diagrams, 
the folded tableaux naturally capture the allowed sets of pivot columns. 

\begin{figure}[htbp]
  \centering
  \includegraphics[height=3.2cm]{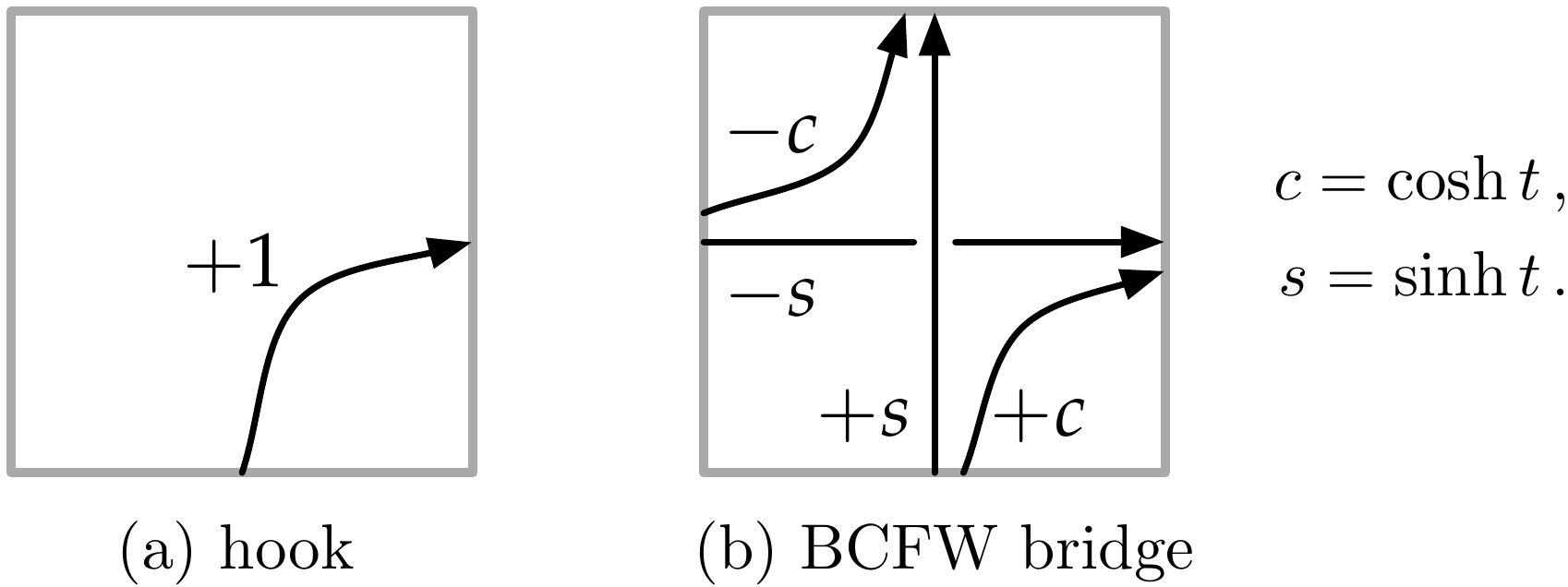}
  \caption{Rules for reading off matrix elements from a tableau}
  \label{fig:tableau-box-rule}
\end{figure}

\begin{figure}[htbp]
  \centering
  \includegraphics[height=4.2cm]{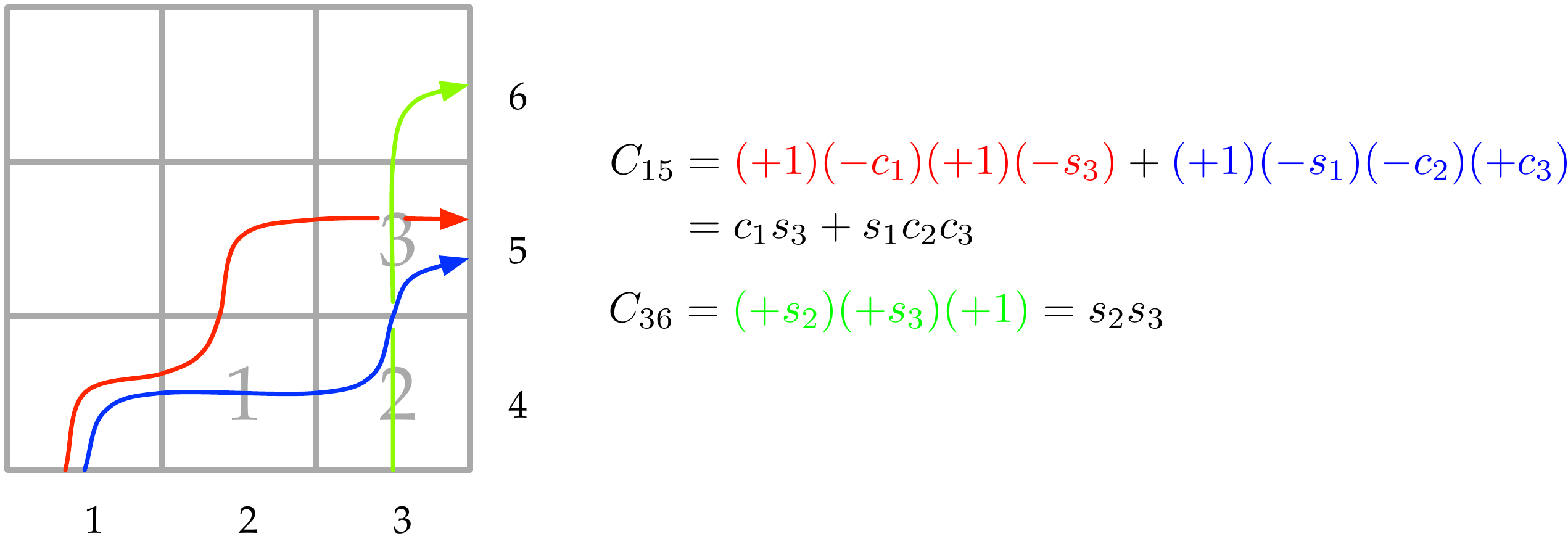}
  \caption{Reading off matrix elements from a tableau: an example}
  \label{fig:tableau-rule-sample}
\end{figure}

The matrix elements of non-pivot columns are determined by 
adopting the conversion rule introduced in Sec.~\ref{sec:conversion}
and modifying it slightly to fit into the tableaux notation. 
Two modifications are needed. First, since the definition of positivity requires a specific branch \eqref{positive-branch}, we are forced to select the orientation-preserving BCFW bridges: $\s=+1$ in 
Fig.~\ref{fig:sc-rule}(a) and $\s=-1$ in Fig.~\ref{fig:sc-rule}(b). 
Second, the labels for external legs are ordered clockwise in on-shell diagrams but counter-clockwise on OG tableaux. 
Taking these factors into account, we arrive at a remarkably simple final rule, depicted in Fig.~\ref{fig:tableau-box-rule}. Note that 
the two types of bridges in Fig.~\ref{fig:sc-rule} has been unified 
into a single one in Fig.~\ref{fig:tableau-box-rule}. An example of the application of the conversion rule is given in Fig.~\ref{fig:tableau-rule-sample}.

At this point, it is not clear how the positivity of Pl\"ucker coordinate 
is related to the positivity of BCFW variables appearing in the conversion rule. To reveal the connection, we will turn to an equivalent, and often more convenient, way to determine the matrix elements. We will begin with bottom cells whose matrix elements are completely fixed by positivty. We will then successively turn on BCFW bridges by multiplying the $C$-matrix from the right by an SO$(k,k)$ rotation matrix. 

\begin{figure}[htbp]
  \centering
  \includegraphics[height=1.7cm]{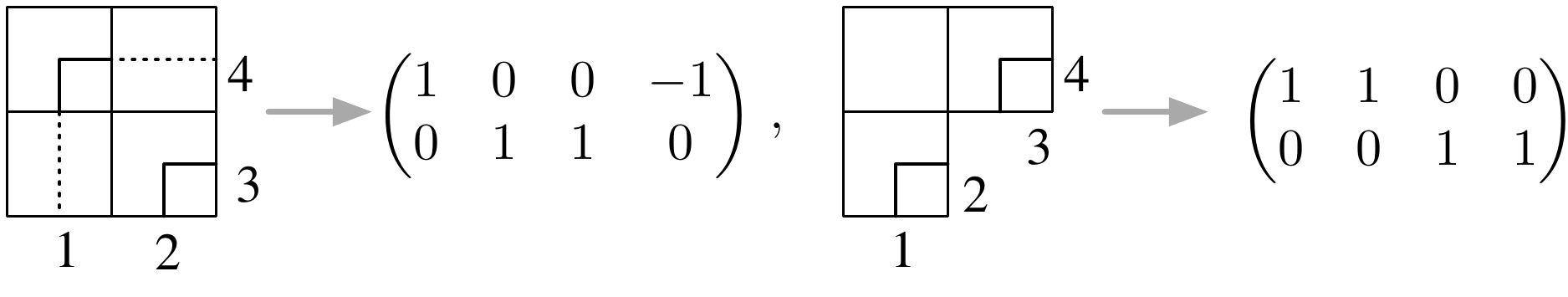}
  \caption{The $C$-matrices for 0-cycles at $k=2$}
  \label{fig:row-echelon-2-bottom}
\end{figure}

\paragraph{Bottom cells}
Each unmarked tableau hosts a unique bottom cell (see Fig.~\ref{fig:tableaux-all-23}). Given an unmarked tableau with pivot columns $\{p_m\}$, let $\{q_m\}$ be the labels for the `sink' columns.  
By construction, $C_{m,p_m}=1$ and $C_{m,i}=0$ for $i\neq p_m, q_m$. The orthogonality constraint requires that $C_{m,q_m} = \pm 1$. 
Positivity determines the sign of $C_{m,q_m}$ uniquely. 
For bottom cells, a minor is non-vanishing if and only if 
it contains either a pivot column $p_m$ or its sink column $q_m$ but not both or neither. We start with $(p_1,p_2,\cdots,p_k)=1$, which is positive by construction. Suppose we replace a pivot column $p_m$ with its sink column $q_m$. 
If the two columns are adjacent ($q_m=p_m+1$), the ordering of the columns in the minor will be preserved, and the minor will remain positive 
if and only if $C_{m,q_m}=+1$. Suppose now $p_m$ and $q_m$ are not adjacent.
Since we are dealing with bottom diagrams with no intersection among lines, the interval 
between $p_m$ and $q_m$ may contain a pair $(p_n, q_n)$ 
for some $n$, but not $p_n$ or $q_n$ separately. 
We can recover the ordering of the minor by shifting 
the column $q_m$ to the right by $(q_m-p_m-1)/2$ steps. 
The resulting ordered minor will become positive 
if and only if we set
\begin{align}
C_{m,q_m} = (-1)^{(q_m-p_m-1)/2}  \,.
\label{bottom-C}
\end{align}
See Fig.~\ref{fig:row-echelon-2-bottom} for 
the explicit form of $C$ matrices for the bottom diagrams at $k=2$. 

\paragraph{BCFW rotation}
Having specified the bottom cells, we can start turning on the BCFW bridges. 
The BCFW bridges act on the $C$-matrix by a right multiplication 
of an SO$(k,k)$ `rotation'. The rotations 
act only on the sink columns and leave the pivot columns intact. 

When there are two or more BCFW bridges, the order of 
the rotation matrices can be determined as follows. 
Let us define the `floor' of a pivot column as their vertical 
distance from the bottom of the OG tableau. 
For example, in Fig.~\ref{fig:tableaux-folding}(b), pivot columns 1 
and 2 reside on the 0th floor, while column 5 resides on the 2nd floor. 

The BCFW bridging begins with those pivot columns on the 0th floor. 
We bridge the two left-most pivot columns, say 1 and 2, 
in the sense that the rotation matrix acts on 
the corresponding sink columns. 
The hook above pivot 1 is lowered from its original location. 
If there are more pivot columns on the 0th floor, say column 3, 
then we bridge pivots 1 and 3. We continue the process until 
the hook above pivot 1 comes down to the 0th floor. 
Finally, we decouple column 1 and elevate other columns 
on the 0th floor to the 1st floor. 

\begin{figure}[htbp]
  \centering
  \includegraphics[height=4.8cm]{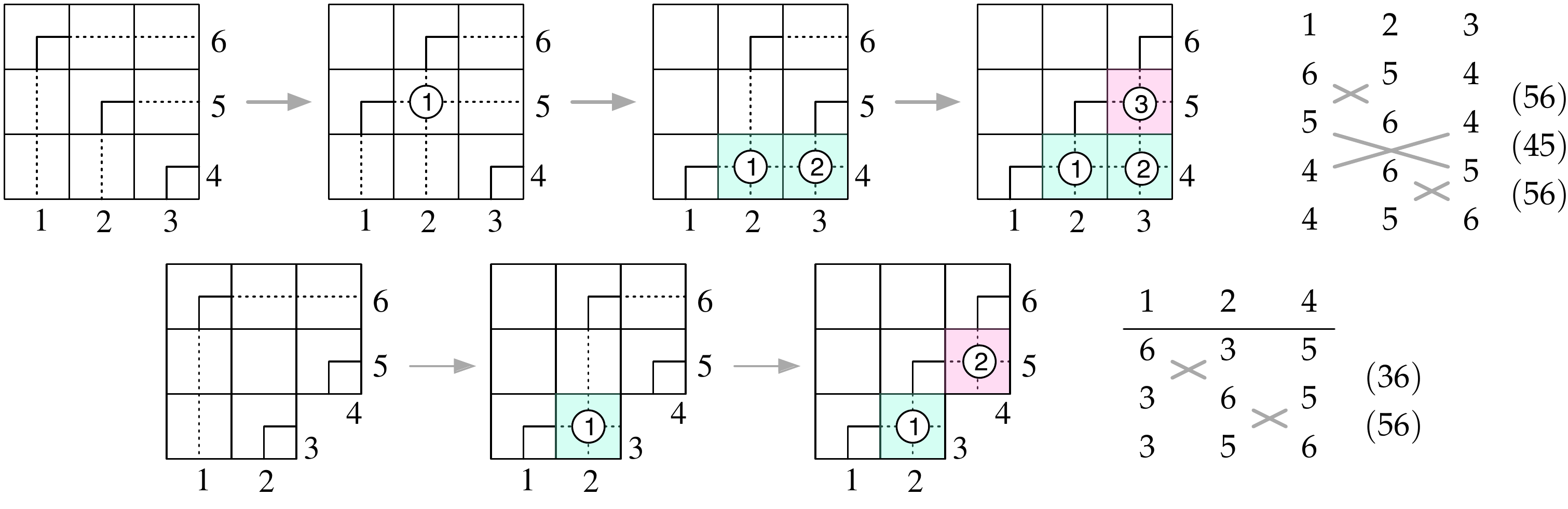}
  \caption{BCFW bridging of the top cell (above) and level 2 cell (below) for $k=3$}
  \label{fig:tableaux-BCFW-3-top}
\end{figure}

\begin{figure}[htbp]
  \centering
  \includegraphics[height=6.5cm]{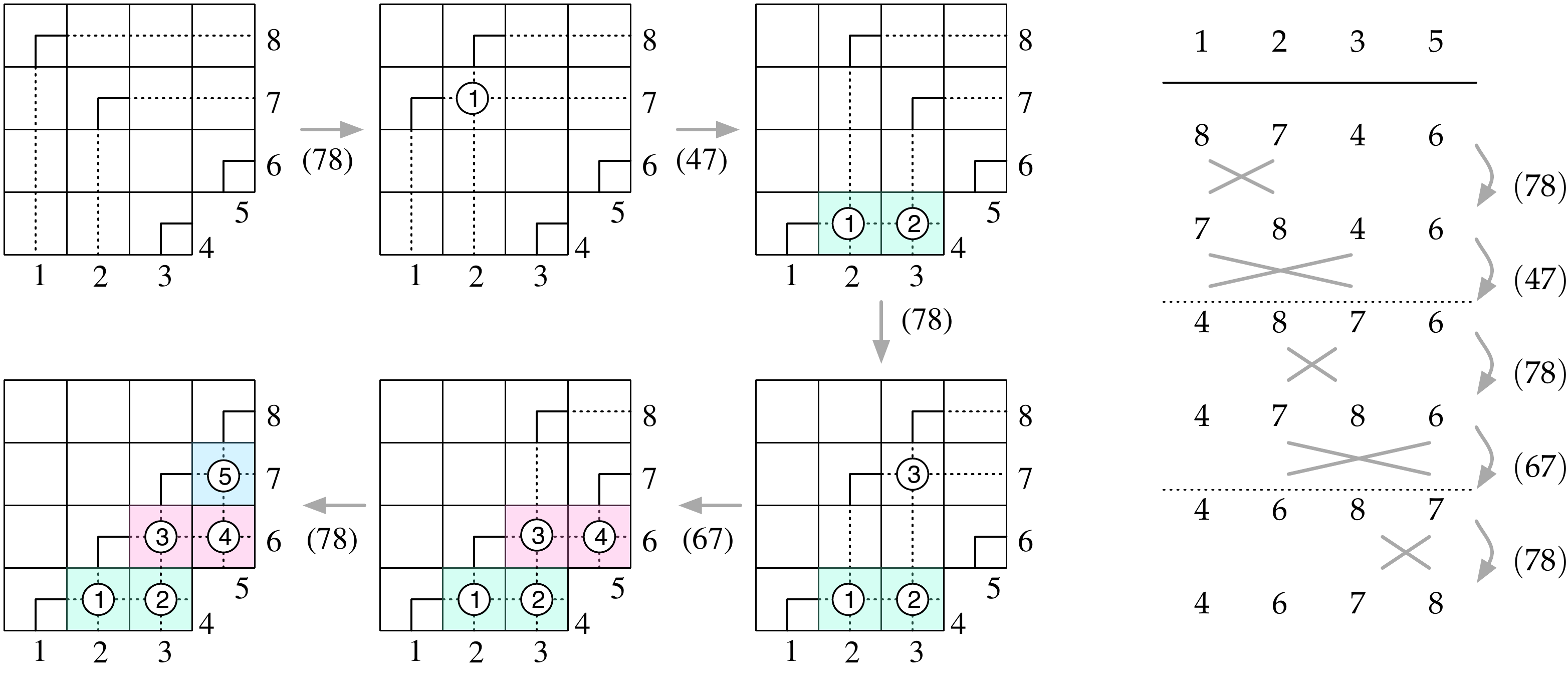}
  \caption{BCFW bridging of the level-5 cell for $k=4$}
  \label{fig:tableaux-BCFW-4-1235}
\end{figure}

We proceed in the same way on the 1st floor including 
those elevated from the 0th floor and those born on the 1st floor. 
The second leftmost column from the 0th floor, 
if exists, becomes the leftmost column of the 1st floor. 
The final result can be summarized in a simple way. 
Given an empty OG tableau, to go from level 0 to the highest level, 
we perform the BCFW bridging such that the order or rotation 
is read off from left to right on the 0th floor, 
and then from left to right on the 1st floor, and so on. 
Two examples are given in Fig.~\ref{fig:tableaux-BCFW-3-top} 
and Fig.~\ref{fig:tableaux-BCFW-4-1235}.

\paragraph{Positivity}

A rotation matrix $R$ for swapping a pair of sink columns $c_i$ and $c_j$ $(i<j)$ is the tensor product of the non-trivial $2\times 2$ block, 
\begin{align}
R_{ii} = R_{jj} = \cosh t \,,
\quad
R_{ij} = R_{ji} = (-1)^{(j-i-1)/2} \sinh t\,,
\label{BCFW-R-sign}
\end{align}
and a $(2k-2) \times (2k-2)$ identity matrix.
The sign factor in \eqref{BCFW-R-sign}, which resembles 
\eqref{bottom-C}, has been inserted to preserve positivity for $t\ge 0$. 
The rotation swaps the sink columns $c_i$ and $c_j$. 
It leads to mixing of the Pl\"ucker coordinates 
$(\cdots i \cdots)$ and $(\cdots j \cdots)$, 
\begin{align}
(\cdots i \cdots) &\quad \rightarrow \quad R_{ii} \; (\cdots i \cdots) + R_{ij} \;(\cdots j \cdots)\\
(\cdots j \cdots) &\quad \rightarrow \quad R_{ji}\; (\cdots i \cdots) + R_{jj} \;(\cdots j \cdots)%\\
\end{align}
When $c_i$ and $c_j$ are adjacent $(j=i+1)$, 
the two minors $(\cdots i \cdots)$ and $(\cdots j \cdots)$ share 
a common ordering. Thus, to ensure positivity of minors after the rotation, we should require that $R_{ij} \ge 0$. When $c_i$ and $c_j$ are not adjacent, the ordering of $(\cdots i \cdots)$ and $(\cdots j \cdots)$ 
differ precisely by the sign factor $(-1)^{(j-i-1)/2}$, 
such that the rotation \eqref{BCFW-R-sign} with the sign factor preserves 
positivity. 
Finally, we note that 
\begin{align}
(\cdots i \cdots j \cdots) \quad \rightarrow \quad &\cosh^2{t}\; (\cdots i \cdots j \cdots) + \sinh^2{t}\; (\cdots j \cdots i \cdots) 
\nn \\
& = (\cosh^2{t}-\sinh^2{t}) (\cdots i \cdots j \cdots)=  (\cdots i \cdots j \cdots)
\,.
\end{align}

We give an explicit form of the $C$-matrix for the two examples shown in Fig.~\ref{fig:tableaux-BCFW-3-top}.
Since the pivot columns form a $(k\times k)$ identity matrix by construction, 
it suffices to present the sink columns. 
For the $k=3$, level 2 cell with pivot columns $\{1,2,4\}$, we have 
\begin{align}
\left(C_3 , C_5 , C_6  \right) &= 
\begin{pmatrix}
0 & 0 & 1 \\
1 & 0 & 0 \\
0 & 1 & 0 \\
\end{pmatrix}
\begin{pmatrix}
c_1 & 0 & -s_1 \\
0 & 1 & 0 \\
-s_1 & 0 & c_1 \\
\end{pmatrix}
\begin{pmatrix}
1 & 0 & 0 \\
0 & c_2 & s_2 \\
0 & s_2 & c_2 \\
\end{pmatrix}
=
\begin{pmatrix}
-s_1 	& c_1 s_2 	& c_1 c_2\\
c_1 	& -s_1 s_2	& -s_1 c_2 \\
0 		& c_2		& s_2 \\
\end{pmatrix} \,,
\label{k=3-top-coord}
\end{align}
where we used the notations $c_i = \cosh t_i$, $s_i = \sinh t_i$. 
For the $k=3$ top-cell with pivot columns $\{1,2,3\}$, we have 
\begin{align}
\left(C_4 , C_5 , C_6  \right) &= 
\begin{pmatrix}
0 & 0 & 1 \\
0 & -1 & 0 \\
1 & 0 & 0 \\
\end{pmatrix}
\begin{pmatrix}
1 & 0 & 0 \\
0 & c_1 & s_1 \\
0 & s_1 & c_1 \\
\end{pmatrix}
\begin{pmatrix}
c_2 & s_2 & 0 \\
s_2 & c_2 & 0 \\
0 & 0 & 1 \\
\end{pmatrix}
\begin{pmatrix}
1 & 0 & 0 \\
0 & c_3 & s_3 \\
0 & s_3 & c_3 \\
\end{pmatrix}
\nn \\
&=
\begin{pmatrix}
s_1 s_2 & s_1 c_2 c_3 +c_1 s_3 & c_1 c_3+ s_1 c_2 s_3 \\
-c_1 s_2 & -c_1 c_2 c_3-s_1 s_3 & -s_1 c_3 -c_1 c_2 s_3 \\
c_2 & s_2 c_3 & s_2 s_3 \\
\end{pmatrix} \,.
\end{align}
It is straightforward to verify that all ordered minors of 
the corresponding $C$-matrices are manifestly non-negative, provided that $t_i \ge 0$.

%\paragraph{Conversion rule revisited}
We have described two ways to construct the $C$-matrix. 
One is to use the conversion rule summarized in Fig.~\ref{fig:tableau-box-rule} and the other is to perform a sequence of BCFW rotations.
The latter exhibits manifest positivity, while the former reveals the connection to on-shell diagrams more clearly.
As an astute reader may have expected, with hindsight, we have adjusted 
the variables in the two approaches such that the they agree without any change of variables. 
Although we have not been able to find a general proof for this agreement, 
we have verified it in all examples up to $k=5$ and expect that it will hold for all $k$. 

\paragraph{Integration measure}
We note in passing that the integration measure is factorized into a $d\log$ form in a way 
similar to that of \cite{ArkaniHamed:2012nw}. Taking the measures 
from the elementary BCFW vertices \eqref{eq:bcfw-4pt-vertex-12},  \eqref{eq:bcfw-4pt-vertex-14} and taking account of Wick rotation, we see that the integration measure for the full Grassmannian integral can be written as 
\begin{align}
%\sum_{ \{\sigma_i\} }
\int \prod_{i} \frac{d t_i}{\sinh{t_i}} = 
%\sum_{ \{\sigma_i\} }
\int \prod_{i} \frac{d z_i}{z_i} = %\sum_{ \{\sigma_i\} }
\int \prod_{i} d \log z_i \qquad 
\left(z_i := \tanh{\frac{t_i}{2}} \right)\,.
\end{align}

%\newpage
\subsection{Polytope}
\label{sec:boundary}

It is known that G$_+(k,n)$ defines a combinatorial polytope also known as `Eulerian poset' \cite{Williams:2007}. 
We will verify one of the requirements for OG$_k$ to be an Eulerian poset. 
%
%\cmt{We will further explore the polytope structure among $\text{OG}_k$ cells, in terms of tools previously developed. We shall begin our dicussion with enumeration of $\text{OG}_k$ cells.}

\paragraph{Eulerian poset}
%We have introduced the $\text{OG}$ tableaux in Sec.~\ref{sub:cell-decomposition}. 
It is straightforward to count the number of cells at each level for arbitrary $k$. The result is most compactly summarized in terms of a generating function, 
\begin{align}
T_k(q) = \sum_{l=0}^{k(k-1)/2} T_{k,l} \; q^l \,.
\label{Tk-def}
\end{align}
$T_{k,l}$ is the number of on-shell diagrams 
with $2k$ external legs and $l$ vertices without any bubble. 
Equivalently, $T_{k,l}$ is the number of OG$_k$ tableaux at level $l$. 
A counting algorithm based on 
the construction of OG tableaux in Sec.~\ref{sub:cell-decomposition} 
can be easily implemented on a computer and generate $T_k(q)$.
The results for $T_k(q)$  
for small values of $k$ are given by 
\begin{align}
&T_2(q) = 2 + q \,,
\nn\\
&T_3(q) = 5 + 6q + 3q^2 + q^3 \,,
\label{Tk-example} \\
&T_4(q) = 14 + 28q+ 28 q^2 + 20 q^3 + 10 q^4 + 4 q^5 + q^6  \,,
\nn \\
&T_5(q) = 42 + 120q+ 180 q^2 + 195 q^3 + 165 q^4 + 117 q^5 + 70 q^6 + 35 q^7 +15 q^8 + 5 q^9 + q^{10} \,.
\nn
\end{align}
After computing $T_k(q)$ up to $k=15$ using our own algorithm, we found that a beautiful closed-form expression for $T_k(q)$ had been known for decades \cite{Riordan:1975d}, 
\footnote{A similar result for G$_+(k,n)$ was obtained in \cite{Williams:2005}.}
\begin{align}
T_{k}(q) = \frac{1}{(1-q)^{k}}\sum_{j=-k}^{k}  (-1)^{j} 
\binom{2k}{k+j} \,q^{j(j-1)/2} \,.
\end{align} 
In special cases, this formula reproduces the simple general features discussed in Sec.~\ref{sub:cell-decomposition}, 
\begin{align}
T_k(1) = \frac{(2k)!}{2^k k!} \,, 
\quad 
T_{k,0} = {\rm C}_k \,,
\quad 
T_{k,k(k-1)/2} = 1 \,.
\label{Tk-general}
\end{align}

\begin{figure}[htbp]
	\centering
	\includegraphics[height=3cm]{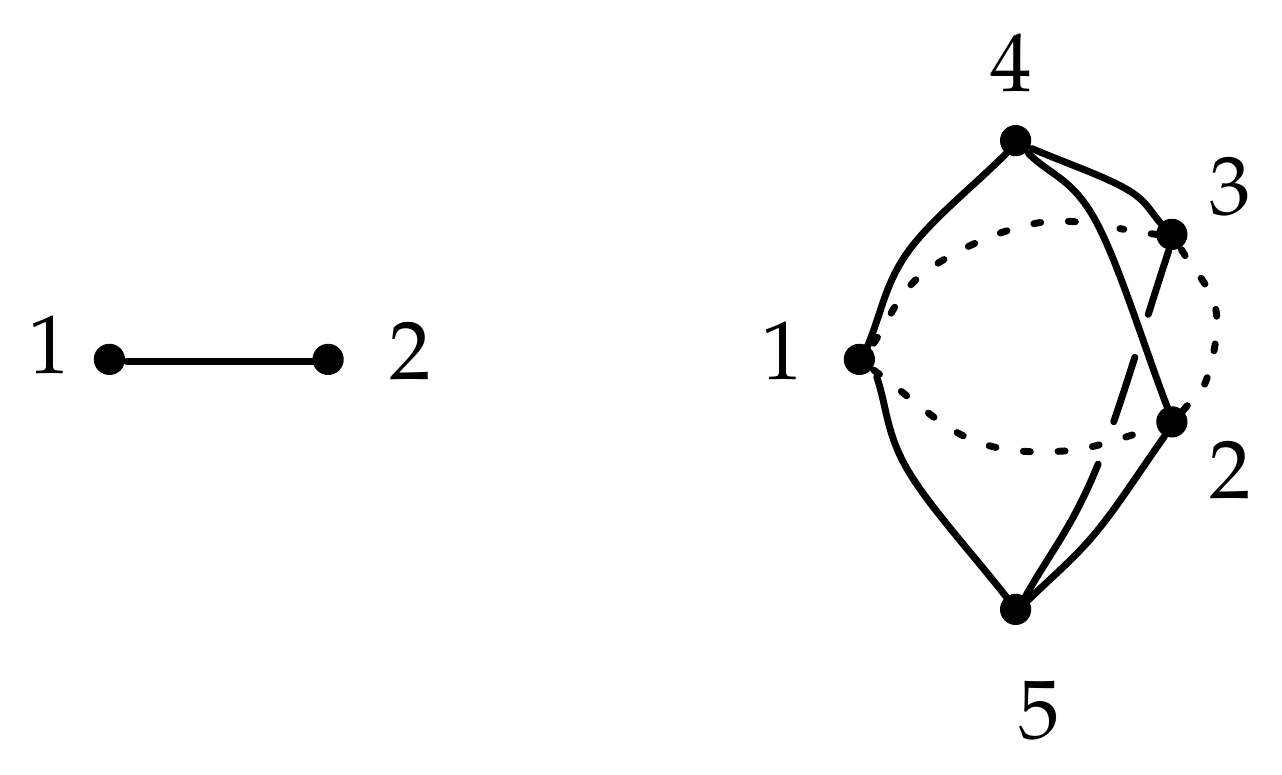}
	\caption{The ball topology of POG for $k=2,3$}
	\label{fig:POG-23}
\end{figure}

Another property of $T_k(q)$ that can be derived 
from \eqref{Tk-general} is that, for any $k$, 
%is the Euler characteristic for the top-cell of positive $\text{OG}_k$ becomes
% In addition, we note two features of $F_k(q)$. First, 
\begin{equation}
	T_k(-1) = \sum_l (-1)^l T_{k,l} = 1 \,.
	\label{eulerian}
\end{equation}
We recognize this as one of the central properties of an Eulerian poset. 
In Sec. \ref{sub:positive-coord}, we assigned a coordinate patch 
of POG$_k$ to each  tableau. In this geometric context, 
$T_k(-1)$ is interpreted as the Euler characteristic of POG$_k$.  
For $k=2,3$ (see Fig.~\ref{fig:POG-23}), \eqref{eulerian} matches 
with the fact that the POG has the topology of a ball. 
It remains to be seen whether the POG is a topological ball for all $k$. 

Using the positroid stratification, we can refine the Eulerian poset structure.\footnote{We thank Yu-tin Huang for bringing 
\cite{Williams:2007} to our attention. See also \cite{Huang:2014xza}.} Consider the $k=2$ and $k=3$ examples in Fig.~\ref{fig:tableaux-all-23}. We can compute an analogue of $T_k(q)$ 
for each chart containing several cells. They turn out to vanish in all charts, 
except for the smallest one with pivot columns $\{1,3,\cdots, 2k-1\}$ which trivially 
gives 1. We checked that the same phenomenon continues up to $k=7$, but have not attempted a proof for all $k$. Geometrically, it indicates that the subspaces of $\text{POG}_k$ are likely to be topological balls. 
For instance, a level 2 cell for $k=3$ shown in Fig.~\ref{fig:tableaux-all-23} is topologically a square with four edges and four vertices.
%
%\textcolor{red}{To gain a better geometric insight for the topology of POG, 
%we turn to our last subject, namely, the boundary operation on on-shell diagrams.}

\begin{figure}[htbp]
	\centering
	\includegraphics[height=1.8cm]{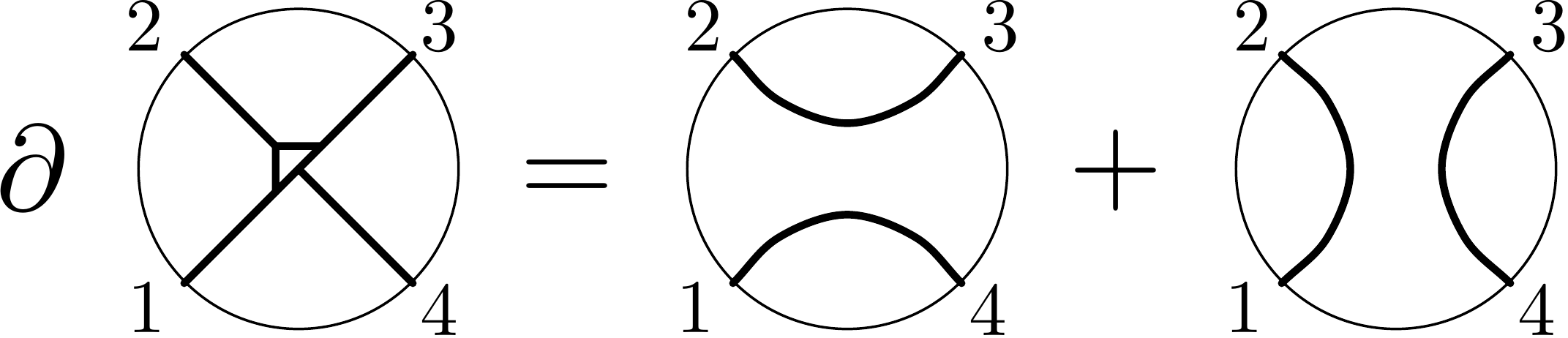}
	\caption{Schematics of the boundary operation.}
	\label{fig:bcfw-boundary}
\end{figure}

\paragraph{Boundary Operation} 
Following \cite{ArkaniHamed:2012nw}, we define the boundary operation $\partial$ acting on on-shell diagrams
such that it resolves each BCFW vertex in two ways shown in Fig.~\ref{fig:bcfw-boundary}. 
The corresponding $\text{OG}_2$ tableaux (see Fig.~\ref{fig:tableaux-all-23}) shows that the first term remains in the same coordinate chart as the original one, $\{1,2\}$, while the second term belongs to $\{1,3\}$. It is a general property of the canonical coordinate system; there always exist some  boundaries that cannot be reached without changing coordinates. 

One possible approach to put every boundary within reach in a single coordinate patch would be to use the cyclic gauge, where the odd-labelled 
columns form an identity matrix and the even-labelled columns 
form an SO$(k)$ rotation matrix. This approach has its own drawbacks. 
First, the cyclic gauge necessarily introduces a BCFW bridge 
with non-adjacent source legs $\{1,3\}$ or $\{2,4\}$. As explained 
in \cite{Huang:2013owa}, this type of BCFW bridge is substantially more 
complicated than the ones used in this paper. 
Second, positivity imposes coupled, non-linear relations among the angle variables of SO$(k)$ rotation, in contrast to 
the simple $t_i\ge 0$ conditions in this paper. For these reasons, we will stay within the canonical coordinate system 
and look for an alternative way to reach all boundaries.

\begin{figure}[htbp]
	\centering
	\includegraphics[height=2.5cm]{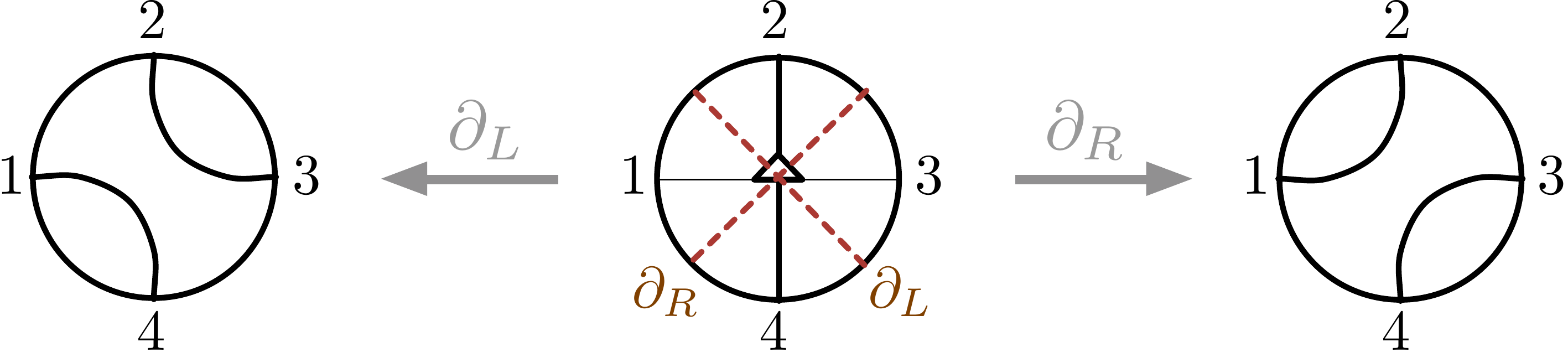}
	\caption{Boundary operation.}
	\label{fig:bcfw-boundary-leftright}
\end{figure}

It is convenient to separate the boundary operation $\partial$ into $\partial_L$ and $\partial_R$, according to
the orientation of the resolved diagrams (see Fig.~\ref{fig:bcfw-boundary-leftright}).

\begin{figure}[htpb]
  \centering
  \includegraphics[height=8cm]{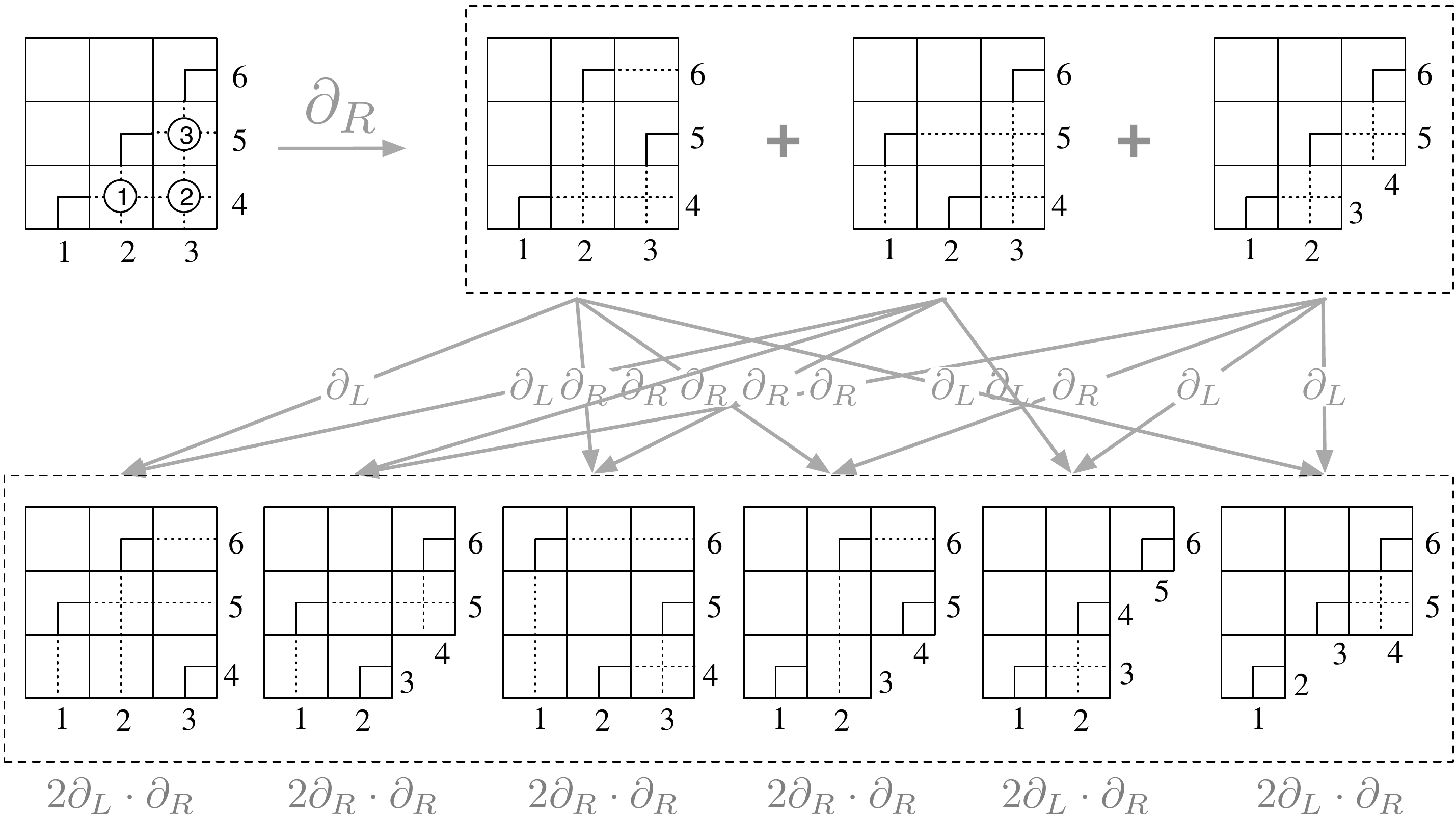}
  \caption{Boundary Operation $\partial^2 = 0 \,(\text{mod } 2)$}
  \label{fig:boundary-k3-top-cell}
\end{figure}

\noindent
In \cite{ArkaniHamed:2012nw}, 
it was shown that $\partial^2 = 0\,\,(\text{mod } 2)$ holds 
for G$_+(k,n)$ and conjectured that the (mod 2) restriction could be dropped if suitable signs are attached to each on-shell diagram. 
\footnote{This conjecture for G$_+(k,n)$ is in fact known. It was shown in \cite{Knutson:2011} that the poset for G$_+(k,n)$ is a subposet of `Bruhat order, 
and it is a classical result that the statement of the conjecture holds for Bruhat order. We thank T. Lam for explaining this point to us.}
Here, we will outline a similar argument for POG using 
an example without attempting a general proof. Applying $\partial$ to the $k=3$ top-cell, we first observe that $\partial_L$ results in bubble configurations, which we will discard by hand. Fig.~\ref{fig:boundary-k3-top-cell} shows not only that each diagram has two incoming arrows, implying $\partial^2 = 0\,\,(\text{mod } 2)$, but also that $\partial_R^2 = 0\,\,(\text{mod } 2) = \partial_L\cdot\partial_R$. In general, we have
\begin{align}
\partial_L\cdot\partial_R + \partial_R\cdot\partial_L = 0\,\,(\text{mod } 2)  = \partial_R^2 = \partial_L^2.
\end{align}

\paragraph{Boundaries of a top-cell} The top-cell at each $k$ 
has precisely $k$ boundaries as can be seen in \eqref{Tk-example}. We will identify them with the vanishing loci of the $k$ independent consecutive minors that appear in the original OG integral \eqref{grass0}. 

We begin with $k=3$ (see Fig.~\ref{fig:boundary-k3-top-cell}). In our coordinate system \eqref{k=3-top-coord}, the three independent consecutive minors are
\begin{align}
(123) = 1\,, 
\quad 
(234) = \sinh{t_1}\sinh{t_2}\,,
\quad
(345) = \sinh{t_2}\sinh{t_3} \,.
\end{align}
We can easily obtain these minors without constructing the whole $C$-matrix. The only relevant matrix elements for $(234)$ and $(345)$ are $c_{14}$ and $c_{36}$, respectively, which can be read off from the paths $1\rightarrow 4$ and $3\rightarrow 6$, respectively. These paths do not involve a turn so that each vertex contributes a $\sinh t_i$. Finally, the construction of POG guarantees that the overall sign should be positive.

A similar argument can be applied to consecutive minors for higher $k$. We simply summarize the results. First, $(123\cdots k)=1=(k+1,\cdots,2k)$ by construction. For other consecutive minors, we can always draw a rectangle within the OG tableau whose right/bottom edges correspond to the sink/pivot columns participating in the minor. If we collect all BCFW variables inside the rectangle, the product of $\sinh t$ factors produces the correct result for the minor. For example, for a general $k$,
\begin{align}
(234 \cdots ,k+1) = \prod_{i=1}^{k-1} \sinh{t_i}, \hspace{0.3cm} (345 \cdots k+2) = \prod_{i=2}^{2k-3} \sinh{t_i}, \hspace{0.3cm} \text{and so on.}
\label{minor-rect}
\end{align}

One may try to reach the boundaries of the top-cell by turning off some $t_i$. However, sometimes it forces two or more consecutive minors to vanish at the same time, leaving a bubble configuration. Among the $k(k-1)/2$ BCFW variables, only $(k-1)$ of them can be safely turned off without 
generating a bubble. It is easy to see that they correspond to the $(k-1)$ vertices along the diagonal of the $\text{OG}_k$ tableau.

By definition of the top-cell, the canonical coordinate does not allow $(123\cdots k)$ to vanish. 
Recall from Sec.~\ref{sub:POG} that, for $k=2$, 
the boundary with `(12)=0' was reached by taking $t\rightarrow \infty$ 
in the canonical coordinates and performing a gauge transformation. 
A similar change of variable, which leads to a different coordinate patch, 
can reach the last boundary of a top-cell.  
In our prescription for computing the minors using rectangles, 
it is clear that all but $(123\cdots k)$ contains $\sinh t_{k-1}$ coming 
from the lower right corner of the OG tableau for the top-cell. 
Taking $t_{k-1} \rightarrow \infty$ makes all consecutive minors except $(123\cdots k)$ diverge. Since only the ratios between minors are gauge invariant, we may divide all minors by $\sinh{t_{k-1}}$ so that 
$(123 \cdots k)$ converges to zero and all other minors remain finite. 
This configuration is equal to the $\text{OG}_k$ tableau whose the rightmost-bottom box is removed. Going back to the $k=3$ example, setting $t_2 \rightarrow \infty$ and dividing all minors by $\sinh t_2$ gives
\begin{align}
(123) = 0, \hspace{0.3cm}
(234) = \sinh{t_1}, \hspace{0.3cm}
(345) = \sinh{t_3}.
\end{align}
This agrees with the result obtained from the $(13)(25)(46)$ tableau
in Fig.~\ref{fig:tableaux-BCFW-3-top}.

In summary, we have observed that $\text{POG}_k$ is likely to form an Eulerian poset and to have a ball topolgy. As a partial attempt to state and prove these conjectures rigorously, we took an initial step to define the notion of boundary operation satifying $\partial^2 = 0$. Such $\partial$ would be naturally identified with the boundary operation of the usual simplicial homology on the geometric side. We verified that $\partial^2 = 0 \text{ (mod 2)}$ works for POG$_k$ up to $k=3$ and is likely to generalize straightforwardly for higher $k$. But, dropping the `modulo 2' restriction seems to be a difficult task. 
We hope to revisit these problems in a future work.

\newpage
%\vskip 1cm 
\acknowledgments

We are grateful to Saebyeok Jeong and Jihye Seo for 
collaboration at early stages of this work. 
We are also grateful to Yu-tin Huang for many enlightening discussions, helpful comments on the manuscript, and sharing a preliminary draft of \cite{Huang:2014xza} prior to publication. 
We thank Lauren Williams for pointing out errors in Sec.~\ref{sec:boundary} of the original version of this paper, 
and Thomas Lam for sharing the preprint \cite{Lam} prior to publication and for many valuable comments. 
The work of JK is supported in part by the BK21 program of the Ministry of Education, Science and Technology of Korea, and the National Research Foundation of Korea (NRF) Grants 2010-0007512, 2012R1A1A2042474 and 2005-0049409 through the Center for Quantum Spacetime (CQUeST) of Sogang University.
The work of SL is supported in part by the National Research Foundation of Korea (NRF) Grants 2012R1A1B3001085 and 2012R1A2A2A02046739.

\end{document}